\documentclass[12pt,preprint]{aastex}

\usepackage{graphics,graphicx,epsf,natbib,lscape}
\usepackage{multirow}
\usepackage{longtable}

\shorttitle{Abundances in X-Ray Early-Type Galaxies}
\shortauthors{Ji et al.}

\begin{document}

\title{Elemental Abundances in the X-Ray Gas of Early-Type Galaxies with
{\it XMM\/} and {\it Chandra\/} Observations}

\author{Jun Ji, Jimmy A. Irwin, Alex Athey, Joel N. Bregman, and
Edward J. Lloyd-Davies}

\affil{Department of Astronomy, University of Michigan, Ann Arbor, MI 48109}
\email{jijun@umich.edu, jairwin@umich.edu, athey@arlut.utexas.edu, jbregman@umich.edu, radix@freeshell.org}

\begin{abstract}

The source of hot gas in elliptical galaxies is thought to be due to 
stellar mass loss, with contributions from supernova events and 
possibly from infall from a surrounding environment.  This picture 
predicts supersolar values for the metallicity of the gas toward the
inner part of the galaxy, which can be tested by measuring the gas
phase abundances.  We use high-quality data for 10 nearby early-type
galaxy from {\it XMM-Newton\/}, featuring both the EPIC and the Reflection 
Grating Spectrometer, where the strongest emission lines are detected
with little blending; some {\it Chandra\/} data are also used.  We find excellent
consistency in the elemental abundances between the different {\it XMM\/} instruments
and good consistency with {\it Chandra\/}.  Differences in abundances with aperture
size and model complexity are examined, but large differences rarely occur.
For a two-temperature thermal model plus a point source contribution, the 
median Fe and O abundances are 0.86 and 0.44 of the Solar value, while Si
and Mg abundances are similar to that for Fe.  This is similar to stellar
abundances for these galaxies but supernovae were expected to enhance the
gas phase abundances considerably, which is not observed.

\end{abstract}


\section{Introduction}

Early-type galaxies possess an interstellar medium that is dominated by hot
gas (3-10$\times $ 10$^6$ K), although the mass of gas can vary widely between
systems.  The origin of the hot gas is not entirely a settled issue, but it is
probably the result of mass loss from stars within the galaxy as well as
infall onto the galaxy, especially when it lies in a galaxy group.  The
abundances of this gas reflect its history and can potentially inform us as
to the number of Type I and Type II supernova that must have been
present.  There have been a number of surprises in the abundance
measurements, such as that the abundance is lower than initially predicted
for gas shed from stars and enhanced by supernovae (e.g.,
\citealt{arim97}).  Also, the values for the gas abundances have varied
considerably, for different models applied to the same galaxy, and
between galaxies, so a uniform picture has been slow to emerge.

There are a variety of issues that face investigators when determining
abundances within early-type galaxies (or other systems with thermal
gas).  There is the problem of instrumental calibration, which can be
notoriously difficult, despite dedicated efforts by the scientific staff.  
It is often several years after launch of a mission before most of the important
calibration issues are understood.  As the calibrations for {\it XMM\/} and
{\it Chandra\/} have matured, this is a good time to examine the spectra of
similar objects and compare the results.  

Another issue is that the derived metallicities are sensitive to the number
of spectral components used in a model.  This was pointed out by \citet{trinc94}, 
among others, who showed that when a single-temperature thermal model
was applied to the luminous emission from an elliptical galaxy, the
derived metallicity was significantly lower than when a two-temperature
model was used.  Often, the $\chi$$^2$ is acceptable for both models, so without
further information, it is difficult to identify the correct model.  The
problem is one of resolution, both spatially and spectrally.  On one hand,
the temperature can vary within a galaxy, so by analyzing projected
emission, there are multiple temperature components.  Also, there
are point sources, mainly due to low-mass X-ray binaries, and these
provide a hard continuum that must be accounted for.  Even with
{\it Chandra\/}, not all of the individual point sources can be excluded, although their
collective spectra are fairly constant from galaxy to galaxy
\citep{irwin03}, so modeling of this component is tractable.  Spectrally,
one could identify the need for various spectral components if it were possible to
measure lines of various ionization states for the same element.  ROSAT
did not have sufficient spectral resolution nor bandpass coverage to constrain many of the
important parameters nor did it have the spatial resolution to remove
point sources.  The {\it ASCA\/} satellite could measure the high energy
contribution from the X-ray binaries, but had very poor spatial resolution
and there were calibration issues at the important low-energy part of the
detector.

Some of these issues are resolved by using {\it Chandra\/} and {\it XMM\/}
observations.  The {\it Chandra\/} data have excellent spatial resolution, so
most point sources can be excluded, and the combination of spectral
resolution and calibration is superior to its predecessors.  In comparison,
{\it XMM\/} has poorer spatial resolution but more collecting area and a
relatively high dispersion grating spectrum is obtained for all on-axis
targets.  

There is a range of results that seems puzzling, as both high and
low metallicities are found in optically similar galaxies.  Individual {\it XMM\/}
observations show a similar range of behavior, with subsolar abundances
(referenced to \citealt{ande89}), as in NGC 6251 \citep{samb04}, NGC
3585, 4494 and 5322 \citep{osul04}, near-solar values, as in NGC 4649
\citep{rand06}, through the supersolar values seen in NGC 507
\citep{kim04}.  Similarly, individual {\it Chandra\/} obervations of galaxies can
be of near solar metallicity, such as in NGC 1316 \citep{kim03} or NGC
4649 \citep{rand04}, but other galaxies can show quite low abundances, such as
NGC 1291 \citep{irwin02}.  Two surveys of galaxies with {\it Chandra\/} also
show a range metallicities, but with no meaningful correlation between
the stellar metallicities and the metallicities of the X-ray emitting gas
\citep{athey07, hump06}.   

If there is a consistent trend, it is with the
ratios of some of the elements, such as in the {\it XMM\/} observation of NGC
4649, where the light elements (N, C, O, Ca; this is dominated by O) are
0.22 of the Solar value, Fe is 0.7 Solar, while the combination of Ne, Mg,
Si, S, Ar, Al, Na, and Ni is 1.7 times the Solar value \citep{rand06}. 
These results of O being low relative to Fe are very common, as are the
relatively higher values of Mg and Si.  Such ratios may be in conflict with
expectations from supernovae \citep{hump06}.

The above studies either have been in-depth investigations of a single
galaxy with a single instrument (often with different models and
approaches), or a study of several galaxies with a single instrument, but
there has not been a comparison of results across instruments.  One of
our objectives is to analyze several galaxies using both {\it Chandra\/} and
{\it XMM\/} data and applying the same models to both data sets.  This helps to
address whether there are serious calibration differences between
instruments that might yield misleading results.  Also, there has been little
use of the Reflection Grating Spectrograph (RGS) data from {\it XMM\/}, which 
has the highest spectral resolution
and where individual strong spectral lines can be detected and identified. 
With advances in how to model extended sources observed with the RGS,
these spectra become extremely useful and we present these for several
bright sources.  Finally, we examine the differences in the abundances that
result when different instrumental backgrounds are adopted.

\section{Sample Selection and Data Reduction}

\subsection{Sample Selection}\label{sec:sample}

The sample is based on the work of \citet{brown98}, who obtained a 
complete sample of 34 early-type galaxies which were optically selected and 
flux-limited. We selected 10 sources in their sample which were observed 
by {\it XMM-Newton\/}, and are bright enough to use both of the data of the European Photon 
Imaging Camera (EPIC) and the Reflection Grating Spectrometer (RGS). 
This is mainly determined by the quality of the RGS spectra, which were 
included in our sample by presenting enough emission line features in the 
preview spectra.  Several of these galaxies also have high-quality 
{\it Chandra\/} data, which were used to determine abundances as well.
The observations that were used in this paper are given in Table~\ref{tab:obs}.


\subsection{{\it Chandra\/} ACIS Data Processing}

{\it Chandra\/} observations of NGC~4649, NGC~4472, and NGC~1399 were obtained 
from the HEASARC online archive (http://heasarc.gsfc.nasa.gov/docs/archive.html).
The data for each observation were processed in a uniform manner following 
the {\it Chandra\/} data reduction threads using {\sc ciao} v3.4. 
Times of high background were removed from the data, and all images created 
were corrected for exposure and vignetting. 
Point sources were identified and subsequently removed using ``Mexican-Hat" wavelet
detection routine {\sc wavdetect} in {\sc ciao} on the 0.3--6.0 keV band image.                                                                                     
For each galaxy, we extracted spectra from a region 1$^{\prime}$ in diameter
centered on the nucleus of the galaxy (the same region as was used with
{\it XMM-Newton\/} data for these galaxies). Background spectra were extracted from a
region near the edge of the S3 chip for each galaxy. While the chosen background
region may contain extended emission from the galaxy (which typically fills
the entire S3 chip), it only represents $\sim$1\% of the total emission in
the 1$^{\prime}$ diameter aperture, so precise treatment of the background is
unnecessary. The spectra were grouped such that each spectral bin contained 
at least 25 counts.

\subsection{{\it XMM-Newton\/}  EPIC data reduction}\label{sec:sample}

The EPIC and RGS data of the 10 sources were processed with standard 
procedures using {\it XMM-Newton\/} Science Analysis System (SAS, Version 7.0.0).  

For the EPIC data, high background periods (counts rate $>$0.35 cnt s$^{-1}$ for 
MOS CCDs and $>$1 cnt s$^{-1}$ for PN CCD) were removed by examining light curves of 
events with photon energy greater than 10 keV and $PATTERN=0$, which 
is very sensitive to the X-ray flares. Photon events with $PATTERN \leq 12$ were 
retained for the MOS cameras while $PATTERN \leq 4$ were retained for the PN camera. We chose 
the photon events in the energy range 0.3-7.0 keV for the MOS and 0.4-7.0 keV 
for the PN to eliminate the calibration uncertainty at the lower energy band 
($<$0.3 keV for MOS and $<$0.4 keV for PN) and the uncertainty at the high 
energy band ($>$7.0 keV) due to very low photon counts. 

Two different source regions were extracted from the cleaned event lists.
The regions are 1$\arcmin$ and 4$\arcmin$ in diameter, centered at the 
peak of the source X-ray emission, which is also the optical center of the galaxy. 
The reason for this is that the smaller region will be more uniform than
the larger region, when radial gradients exist in the gas, such as temperture
gradient.  The larger region is representative of the average properties of the
entire galaxy, so this analysis might be useful for comparisons with more distant 
galaxies. The larger diameter region will sample a broader variation of gas 
properties, such as in temperature and abundance, so by comparing the spectral
fitting results from the two regions, we can assess the implied variations.

We subtract the background regions on the same observation fields, which 
are a few arc-minutes away from the central sources and contain little 
hot gas emission from the central sources; point sources were excluded. 
The background regions are larger than the source regions to compensate 
the vignetting effect on the edge, and are circular regions with diameters 
of 1.67$\arcmin$ and 5.83$\arcmin$ for source region of 1$\arcmin$ and 4$\arcmin$, 
respectively.  Redistribution matrix files (RMF), which contain the 
instrumental responses, and ancillary response files (ARF), which 
contain the effective area of the detector, were generated with SAS.
The spectra were grouped so that each energy bin contains a minimum 
of 25 counts, appropriate for $\chi^2$ fitting.

\subsection{{\it XMM-Newton\/} RGS data reduction}

For the RGS data, high background periods were screened by examining the 
light curve of CCD9 (excluding the source region), which is the most sensitive 
CCD to be affected by the background flares.  The screening criteria is the 
same as employed by \citet{tamu03}, because we are using the same template 
RGS background.  Events were selected in the energy range of 0.33 -- 2.48 keV 
(5 -- 38 \AA), and the first order spectra of RGS1 and RGS2 were extracted. 

The RGS is a slitless spectrograph with a field of view of 5\arcmin\ in
the cross-dispersion direction and 1\arcdeg\ in dispersion direction 
(covering the whole diameter of MOS field of view).  
The RGS1 and RGS2 are perpendicular to each other and therefore
sample different projected spatial regions of the galaxies.  
The size of the extended emission can be inferred from the their
EPIC images, and we can fit the suface brightness distribution with 
a ``beta" model, $S_x \propto (1 + (r/r_{c})^{2})^{-3\beta +1/2}$, 
where $\beta \sim 0.5$ and r$_c$ $<$ 0.2\arcmin.  Although some of
these galaxies are detectable in X-rays to 5\arcmin\, the flux is 
concentrated.  Based on the surface brightness distribution, we
defined two source extraction regions in the cross-dispersion direction.
For a cross-dispersion width of 1\arcmin\ (to be compared with 1\arcmin\ 
diameter regions), 90\% of the flux is extracted,
while in our larger extraction width of 3.5\arcmin, 99\% of the flux is extracted.
These extraction regions are somewhat different than the circular apertures
used to analyze the EPIC data, which is a consideration in the spectral fitting process.
If there is a gradient in spectral properties, the narrower extraction width
will suffer less spectral blending.  The advantage of the spectrum with the
larger extraction width is higher S/N, provided that the source remains
brighter than the background.

When extracting a spectrum, a background must be removed and the standard
procedure is to use a local background, defined as a region in the same 
CCD, after excluding sources. 
However, some of these galaxies fill the field of view of the RGS CCDs, so there is no
emission-free background region that can be used.
Therefore, we used a background from other observations.  For our observations, we used
the RGS template background \citep{tamu03}, combined with several other blank sky observations.
The background is selected with the same region on RGS CCDs of the template as the source region. 
Another advantage of this method is that the background is better defined than a local background
due to the very long exposure times of the background fields.

Following screening and background subtraction, the RGS spectra were binned so 
that each energy bin contain a minimum of 25 counts, making the data appropriate 
for $\chi^2$ fitting. 
 
\section{Spectral Fitting Techniques and Results}
\subsection{Spectral Fitting Techniques}

\begin{figure}
\epsscale{1.0}
\plotone{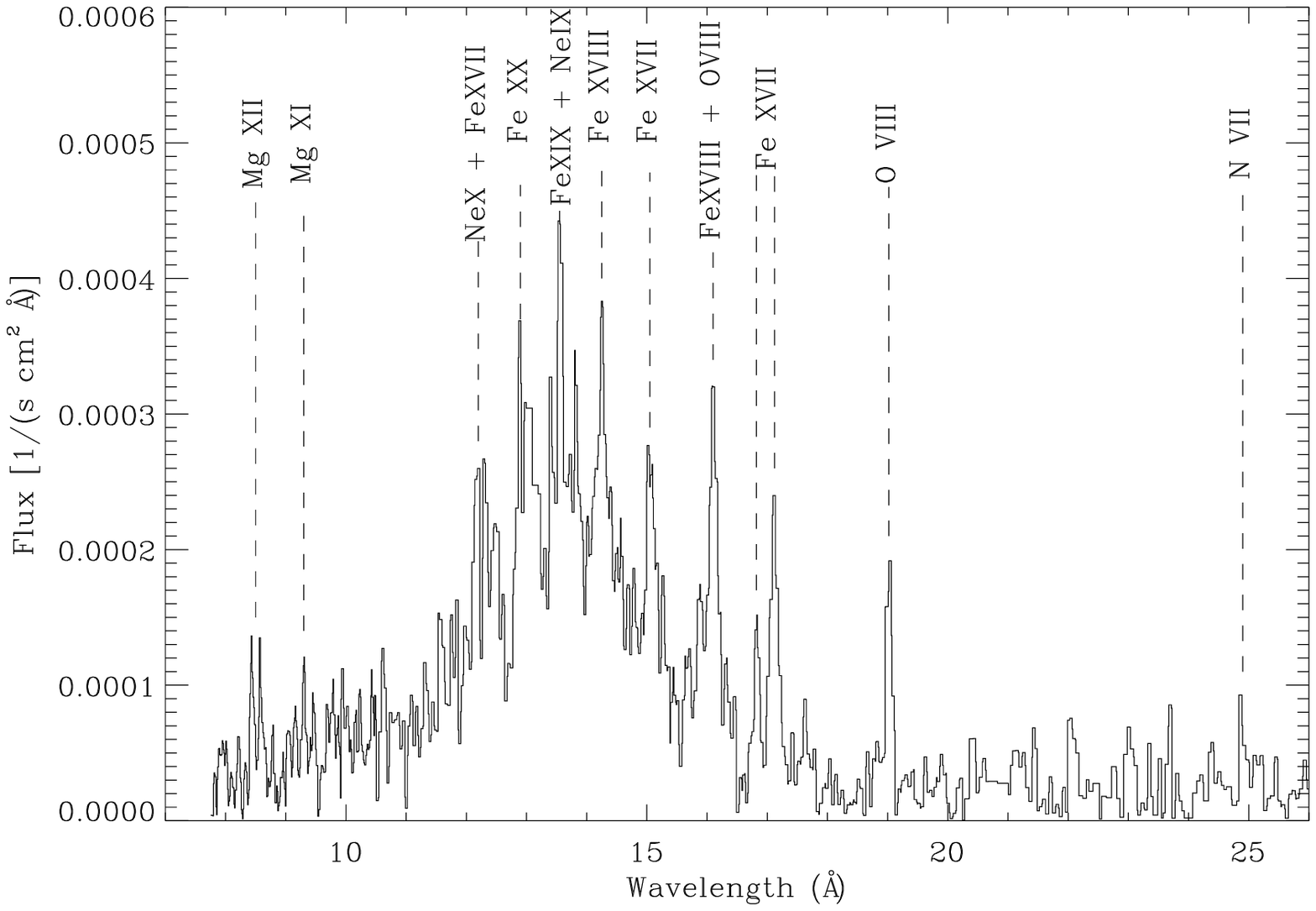}
\caption[RGS spectrum with marked lines]%
        {{\it XMM-Newton\/} full-range RGS spectrum of NGC4472 with marked emission lines from 7 \AA~ to 26 \AA~ band.}
\label{fig:RGS}
\end{figure}

Originally, we anticipated using the RGS spectral fits to yield unambiguous
abundance results, as individual lines are resolved from OVIII (19 \AA ) and highly
ionized Fe, such as FeXVII (15, 17 \AA ) and FeXVIII (16 \AA ) (see Fig.~\ref{fig:RGS}), 
whereas they are badly blended in the EPIC data.  
These RGS data provide clear identification of the line strengths and
the absence of certain important lines, such as the OVII triplet (20.6-21.1 \AA ).
However, the absolute abundance is a ratio of the line fluxes to the
continuum fluxes, and this is less well determined in the RGS data (compared
to the EPIC data) for two reasons.  The first is that the S/N is poorer in the RGS
data due to fewer photons.  The second point is that there is a power-law
continuum contribution from LMXBs, which can be constrained at high energies
from the higher S/N EPIC data.  This power-law component cannot be properly
constrained with the RGS data because of lower S/N and because the instrument
has no response above 2.5 keV, where the power-law is most distinct.  Therefore, we found
it useful to consider the EPIC and RGS spectral fits separately as well as jointly.
All spectral fitting was performed within XSPEC (Version 11.3.2). 

As discussed below, there is good consistency when fitting the EPIC
instruments separately, so we permfom a simultaneous fit of a single model to
the MOS1, MOS2, and the PN data with their normalizations free to fit.
For the thermal emission from thermal gas, we used the Astrophysical Plasma Emission
code (APEC, \citealt{smith01}), which models optically-thin emission from an isothermal plasma
of some abundances ({\it vapec\/}).  The optically-thin assumption should be appropriate
in most cases, although \citet{xu02} demonstrate that optical depth effects 
occur in the strongest \ion{Fe}{17} lines for the galaxy NGC 4636.  
Variations in temperature are accomodated by introducing
two {\it vapec\/} models with different temperatures but same set of abundances.
For the fits, we fix the X-ray absorbing column at the Galactic HI value \citep{dick90},
and we fit for the temperture, the abundances of Fe, O, Si, and Mg, and the normalization.
When the data had sufficiently high S/N, we also fit for Ne, Ni, and S, otherwise, they
were tied to Fe with the ratios given by the Solar values.  All other elemental abundances
were tied to Fe, except for He, which was set at the Solar value.
The abundances are referenced to the Solar values, which are defined here by the
values of \citet{grev98}.  In many earlier works, the Solar abundances of \citet{ande89}
were used, which have higher concentrations of O and Fe by 26\% and 48\%.
Using the newer Solar abunandances, the relative abunances are raised. 

When fitting for the power-law component, most point sources are unresolved,
unlike the situation for the {\it Chandra\/} observations.  
Based on {\it Chandra\/} observations of point sources in early-type galaxies, the
summed x-ray point source spectra can be represented by a power law model 
with an average photon index $\Gamma$ of 1.56 \citep{irwin02}. This component dominates mainly 
the hard x-ray band ($>$ 2 keV), but also contributes some emission to the soft x-ray band ($<$ 2 keV). 
In the models, the contribution from this model is implemented by including a
power-law component, with a fixed photon index of $\Gamma$ = 1.56 and a free normalization; 
this component adds one free parameter.  From our tests, we find nearly the
same value for the power-law normalization when we only consider data above 3 keV (and 
no thermal component) or when we fit the entire energy range of the spectrum,
using a power-law and a {\it vapec\/} component.

The procedure for fitting the {\it Chandra\/} spectra are similar except that
most of the point sources are excluded from the spectral fits.  Still, there
remains the contribution from unresolved point sources, so we included a
power-law component, with fixed $\Gamma$ to 1.56, along with the themal component
({\it vapec\/}) and a Galactic absorption term ({\it wabs\/}).

For the fits to the RGS data, there are a few other considerations, the most important
being the broadening of the lines along the dispersion direction due to the 
extended nature of the emission.  A correction for this can be made if
the spatial distribution of the emission is known, and for this purpose, we
used the EPIC MOS1 image (equivalently, the other images could have been used).
Using this image, the broadening correction is implemented by the task {\it rgsxsrc\/}
within XSPEC.  In the first set of fits to the RGS data, we fix the ratio of the luminosities
from the thermal plasma component ({\it vapec\/}) to the power-law component ({\it powerlaw\/})
at the value given by the fits to the EPIC data.  For the thermal
plasma model, we use the RGS data to fit for the temperature, the
normalization, and the abundances for O and Fe, which are the element abundances 
that are best constrained by the RGS data.  Abundances for Mg and Si
are determined whenever possible, otherwise they are tied to Fe. In addition, lines are
detected from Ne and N for some sources, but the S/N is not sufficient for 
useful determinations, so the abundances of N, Ne, S, and Ni are tied to Fe.

\subsection{Comparison of ACIS-S and EPIC Spectral Fits for Bright Sources}\label{sec:cross-compare}

\begin{figure}
\epsscale{1.0}
\plotone{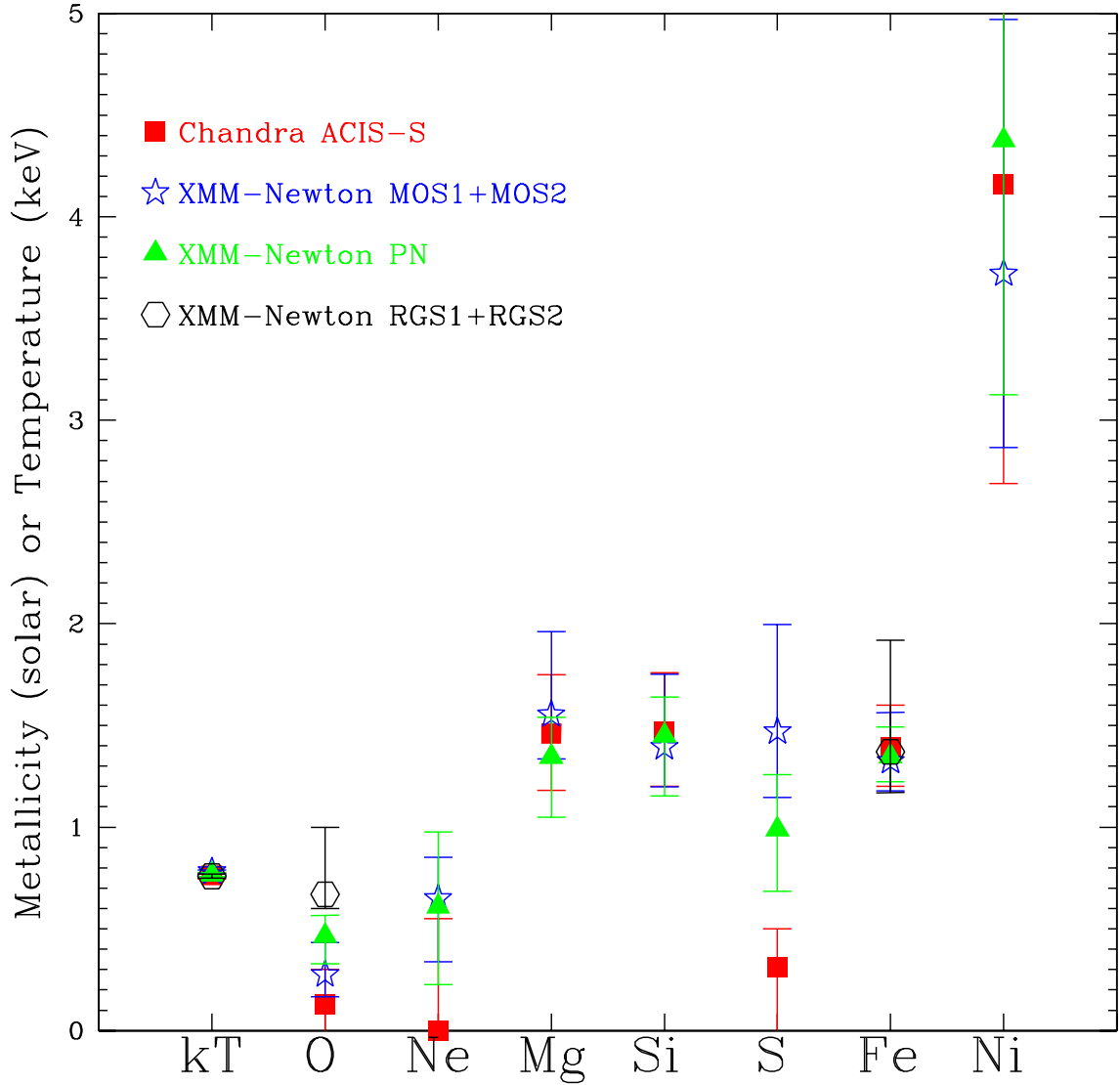}
\caption[cross-comparison of abundance fits]%
        {The temperature and abundances from the {\it XMM-Newton\/} instruments and
        the {\it Chandra\/} ACIS-S for NGC 4649.  A 1\arcmin\ diameter region was used,
        along with a model that includes Galactic absorption, power-law emission
        from binary sources, and a one-temperature thermal model where abundances
        are fitted for the elements shown.  The results from the {\it XMM-Newton\/} instruments
        are in agreement with each other and with the {\it Chandra\/} results, with the 
        exception of O and possibly S.  Elements heavier than Ne are modestly 
        supersolar, while the two lighter metals, O and Ne, have abundances about
        a factor of two lower.}
\label{fig:n4649compare}
\end{figure}

In the comparison between the four imaging devices (see Fig.~\ref{fig:n4649compare}), the results
for the inner 1\arcmin\ diameter of NGC 4649 were fairly consistent, especially
between the different EPIC instruments.  Using the model described above,
best fits were obtained separately for the ACIS-S3 data,
the PN data, the combined MOS1+MOS2 data, and the combined RGS1+RGS2 data.
This galaxy is an important case because the temperature gradient is
small and there is no evidence for a multi-temperature gaseous medium.
We find that the
temperatures were nearly identical for all detectors, but there were differences in
the abundances of O and S between the {\it XMM\/} EPIC instruments and the
{\it Chandra\/} ACIS-S detector.  The S abundance obtained from the ACIS
data is 75\% lower than that derived from the EPIC data, and the 90\%
error bars do not overlap.  However, the two abundance measurement
differ at the 1.8$\sigma$ level, so this is not a major discrepancy.  A similar, but
slightly greater discrepancy exist for the O abundance, where the EPIC
value is about 2$\sigma$ above the ACIS value.  Aside from these differences,
the abundances follow rather closely for the different instruments in that,
relative to Solar values, O is the least abundant element, followed by Ne. 
Relative to oxygen, iron has a significantly higher abundance, with Mg and 
Si a bit higher yet, and then Ni is much higher than the other elements.

\subsection{Comparison of the RGS and EPIC Results}\label{sec:RGScompare}

We considered one-temperature models in both the central 1\arcmin\ (for both EPIC and RGS data) and more
extended 4\arcmin\ diameter regions (for EPIC)/3.5\arcmin\ regions (for RGS), as well as two-temperature models (2T thereafter), 
shown in Tables 2-3.  For the single temperature fits (Table 2), we give 
both the EPIC (MOS + PN) and the RGS results, and aside from the temperature and
abundances, we list the degrees of freedom in the fit (dof), the reduced $\chi^2$,
and the luminosities from the power law component (Power) and from the thermal component 
in the  0.33 -- 2.48 keV band (in units of 10$^{40}$ erg s$^{-1}$).  
The final column is the ratio of the thermal to the power-law luminosities.  
For the fit to the RGS data, we forced this ratio to be the same value as 
for the EPIC fit.  In Table 3, we fit all the EPIC data and some good quality RGS data, 
but with a two-temperature thermal model, so there are two additional columns, for the second temperature,
and for the luminosity from the second thermal component.  The ratio of the 
luminosities is the sum of the two thermal components divided by the power-law
component.  There are two rows per instrument in each table, for the 1\arcmin\ (for both EPIC and RGS) and the
4\arcmin\ (for EPIC)/3.5\arcmin\ (for RGS) diameter regions.  

\begin{figure}
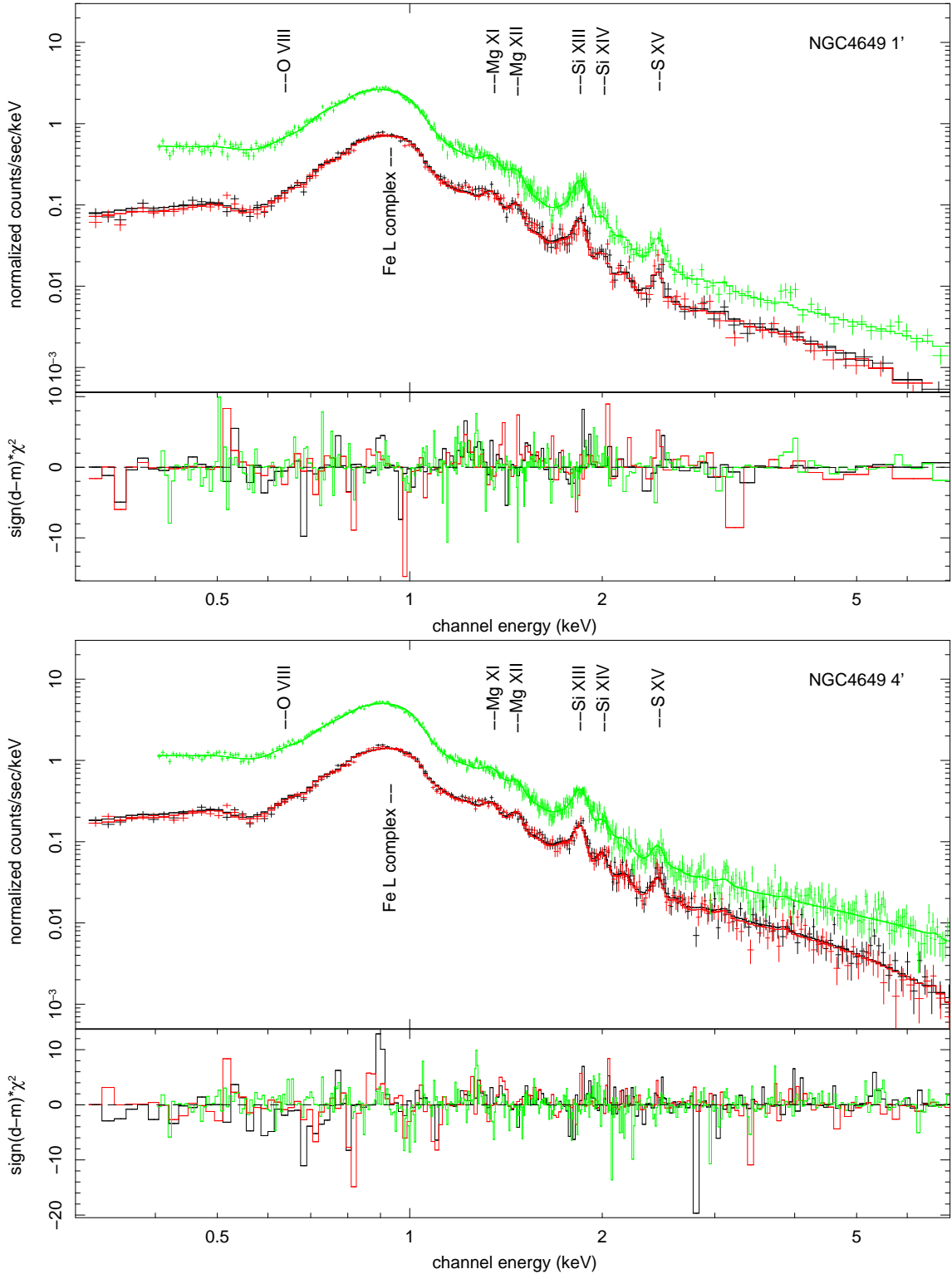

\epsscale{1.0}
\includegraphics[width=4in,angle=-90 ]{f3a.eps}
\includegraphics[width=4in,angle=-90 ]{f3b.eps}
\caption[NGC4649 EPIC]%
        {Example of NGC4649 for {\it XMM-Newton\/} EPIC spectral fitting with marked emission lines. The upper panel is for 1\arcmin\ aperture region while the lower panel is for 4\arcmin\ aperture region. (Green):pn camera; (red): MOS1 camera; (black): MOS2 camera. Note: the emission lines from Fe L complex and O VIII are completely unresolved while emission lines from Mg, Si, and S can be indentified fairly well in CCD spectra.}
\label{fig:n4649fits}
\end{figure}

When fitting a model to the data, only photon statistics define the standard
deviation per point.  In addition, there are systematic uncertainties in the
calibration of the instrument as well as instrumental artifacts, not all of
which are known or excluded from spectral fits.  There is no way of including
accurately such effects in the spectral fitting, but we can identify the
way in which the $\chi^2$ will be affected.  The resulting $\chi^2$ will be 
larger than if there were no systematic effects.  Furthermore, these systematic
effects do not change as the source becomes brighter, but brighter sources have
more photons and therefore lower uncertainties due to photon statistics.
Therefore, systematic errors in valid spectral fits will lead to significantly 
higher reduced $\chi^2$ and fits that formally would be rejected.
This can lead to one rejecting spectral fits to all of the best data sets.
Consequently, we have inspected every spectral fit and there are several
that appear to be good fits but would formally be rejected. In Fig.~\ref{fig:n4649fits} we show such an example of the best fit 
model (2T model) to the EPIC spectra of NGC 4649, which has unacceptable reduced $\chi^2$ values (1.350 for 1\arcmin\ region and 1.417 for 4\arcmin\ region) 
but shows no significant deviations between best-fit model and data. 

For the single-temperature fits within the central 1\arcmin\ (Table~\ref{tab:1T}), 
formally acceptable $\chi^2$ values (not ruled out above the 99\% confidence level)
are found for three objects with fits to the EPIC data (NGC 720, NGC 4552, and NGC 5044)
and two objects with fits to the RGS data (NGC 720 and IC 1459, although NGC 720 has the fewest RGS photons of any source).
Of the sources with a large number of photons, which can be inferred from the number
of degrees of freedom, the fit to NGC 4406, NGC 4649, and NGC 5044 appear to 
be good, while there are clear problems with the fits to NGC 1399, NGC 4636, and NGC 4472.
The galaxy IC 1459 has a strong central point source that makes it difficult
to isolate the thermal gas that we wish to study, so we exclude this from
further discussion (but we gave the 1T fitting results in Table 2 excluding central 13\arcsec\ region in radius).
The fit using the 4\arcmin\ diameter region is worse than when using the 
central 1\arcmin\ region, so we emphasize those 1\arcmin\ region results.

In these one-temperature models, there is quite good agreement in the value 
of the temperature between the EPIC and RGS fits, and there is also good agreement in the Fe and O abundances between the EPIC and RGS fits within 90\% error range 
except for
NGC 720, NGC 3923, NGC 4552, and IC 1459, which have very poor RGS spectra to constrain abundance parameters (Table~\ref{tab:1T}).  Excluding
the galaxies NGC 1399, NGC 4472, NGC 4636, and IC 1459 (see above), the range of the
Fe and O abundances from the EPIC are 0.42-1.57 and 0.26-0.55, respectively
(relative to the Solar values).  The median values for Fe and O are 0.58 and 
0.35 of the Solar values, while the median values from the RGS fits (also excluding NGC 720, NGC 3923, NGC 4552) are
slightly higher, at 0.65 and 0.42.  The ratio of the Fe to O abundance is around
2:1 in this data set.

Of particular interest are three galaxies, NGC 4406, NGC 4649, and NGC 5044,
where the RGS fits appear to be
good and where the abundance uncertainties are not large (despite having 
formally unacceptable values of $\chi^2$, discussed above).  The galaxy 
NGC 4649 was discussed above and here we add that the RGS metallicities
are in good agreement with those derived from the EPIC data.  This is one
of the few galaxies where the Fe metallicity is supersolar, although only
modestly.  The galaxy NGC 5044 has consistent abundances from the EPIC and
RGS data, which show that the Fe abundance is about 0.74 of Solar and the
O abundance is about 0.43 of Solar.  The third galaxy, NGC 4406, has consistent 
O abundances (about 0.36 Solar) between the two {\it XMM\/} instruments, but the Fe abundance 
is higher in the EPIC observations.  This is a complicated galaxy that is 
interacting strongly with its environment \citep{stickel2003}, so perhaps
it is not surprising that there would be disagreement for two instruments
that sample slightly different spatial regions.

When a second thermal component was introduced, the spectral fits generally
improved and the metallicities usually increased. (We fit only six sources with good quality RGS spectra for a second thermal component added. The 
second temperature will be frozen to its EPIC value if it can not be constrained. We list all results but NGC 4649 in Table~\ref{tab:2T}, which shows no convergence for
 the 2T model.) For the same eight galaxies used here (all except NGC 4636 and IC 1459) and for one-temperature fits, the Fe abundance rose
by a median of 30\% while the O abundance rose by 4\%. 
For the EPIC fits to the data within the central 1\arcmin\ (Table~\ref{tab:2T}),
four of the galaxies had formally acceptable fits (NGC 720, NGC 3923,
NGC 4552, and NGC 5044), NGC 4406 is nearly acceptable, and the fits to 
the others appear to be good, with the exception of NGC 4636.  Of the good
and formally acceptable fits (all except NGC 4636 and IC 1459), the range 
in the Fe abundance is 0.44-1.70 Solar and for O it is 0.27-0.76 Solar, with 
medians of 0.86 and 0.44 Solar.  Even the two galaxies excluded from the
analysis had similar values.  The abundances for Mg and Si are indistinguishable
from that of Fe, whereas in the one-temperature fits, they were systematically
higher than the Fe abundance.  In these fits, most of the power is from the
gas phase, typically more than 90\% in the 0.33 -- 2.48 keV band. The RGS fits have Fe and O abundances of 0.33-1.50 Solar and 0.31-1.22 Solar,
 with medians of 0.95 and 0.66 solar, which is consistant in Fe but higher in O.

\begin{figure}
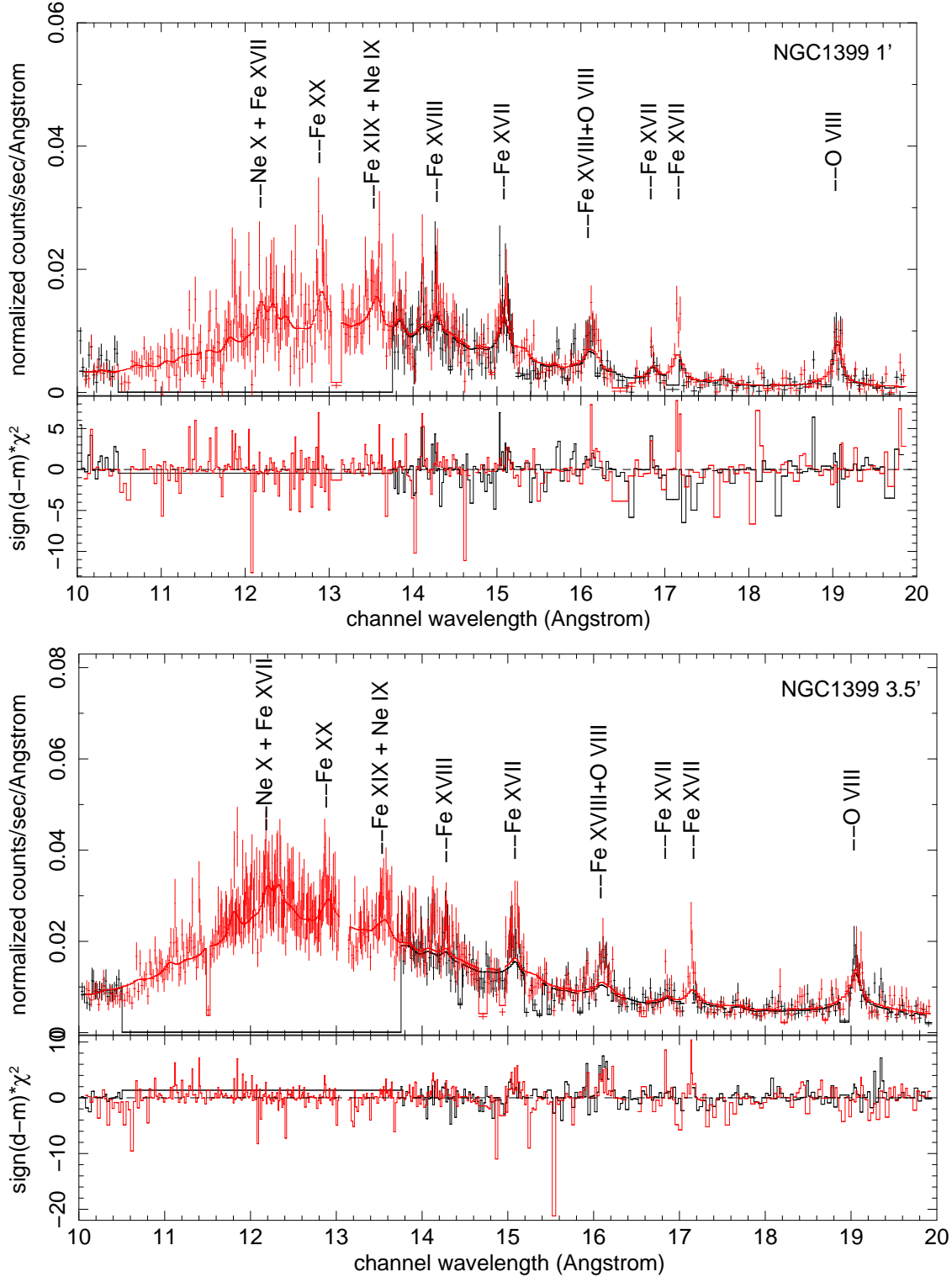

\epsscale{1.0}
\includegraphics[width=4in,angle=-90 ]{f4a.eps}
\includegraphics[width=4in,angle=-90 ]{f4b.eps}
\caption[RGS:n1399]%
        {The 10-20 \AA\ region of the {\it XMM-Newton\/} RGS1 and RGS2 spectra for NGC 1399 with the best-fit models. The upper pannel is for 1\arcmin\ aperture while the bottom one is for 3.5\arcmin\ aperture, with marked resolved emission lines mostly from the Fe L complex and O VIII. Note that the wide gap between 10.5 \AA~ and 13.8 \AA~ is due to failure of RGS1 CCD7, while the narrower gaps at 11.4 \AA, 13.1 \AA, 14.6 \AA, 15.3 \AA, 16.5 \AA, 17.1 \AA, and 18.9 \AA~ are due to gaps between RGS CCDs.}
\label{fig:n1399}
\end{figure}

\begin{figure}
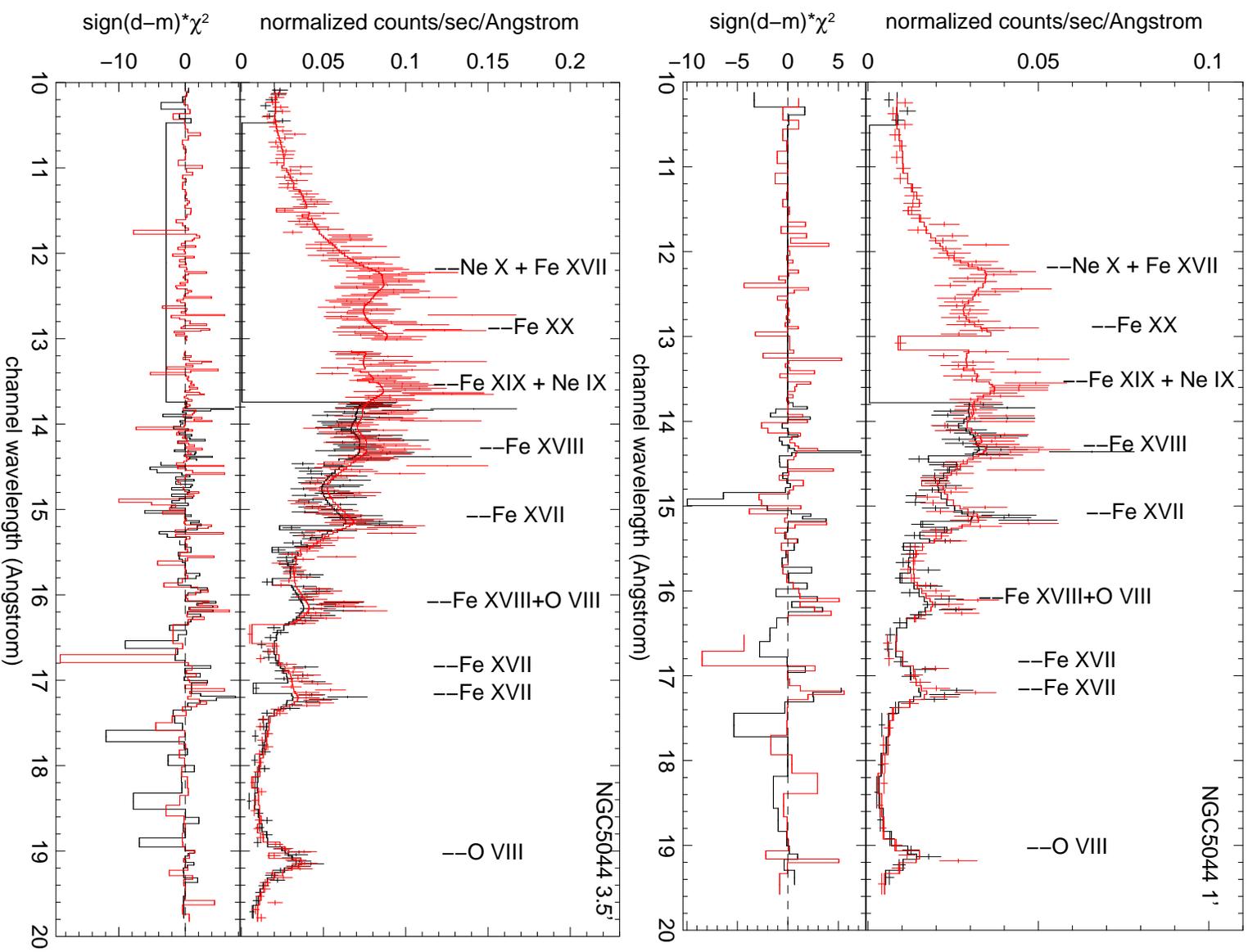

\epsscale{1.0}
\includegraphics[width=4in,angle=-90 ]{f9a.eps}
\includegraphics[width=4in,angle=-90 ]{f9b.eps}
\caption[RGS:n5044]%
        {Same as Fig.~\ref{fig:n1399}, but for NGC 5044.}
\label{fig:n5044}
\end{figure}

In Fig.~\ref{fig:n1399} -~\ref{fig:n5044}, we show zoomed-in RGS spectra with best fit model for the six most bright galaxies with marked individual emission lines. Note that, overall
 the fits are fairly good, but there are some residuals at Fe XVII and Fe XVIII at 15 \AA, 16 \AA, or 17 \AA. Those residuals are not due to improper convolution with large
 image size, since usually the isolated O VIII line at 19 \AA\ is fitted fairly well. They might be due to calibration issues or non-accurate APEC model at those lines.

One important object for comparison is NGC 4636, where \citet{xu02} fit a model
to the RGS data within 2\arcmin\ that included a temperature gradient and optical 
depth effects for \ion{Fe}{17}, which necessitated the use of a $\beta $ model.
Their abundances for Fe and O are 1.29 $\pm$ 0.06 and 0.63 $\pm$ 0.06, relative
to the Solar abundances of \citet{grev98}.  Their ratio of Fe to O is 2 in these 
units.  From our fitting of the RGS data with a single temperature and no opacity
correction, we obtain Fe and O abundances that are about 30\% lower, but with 
the same abundance ratio.  Our somewhat lower abundance for Fe might be 
due to the absence of optical depth corrections. \citet{xu02} found that the optical depth of Fe XVII line at 15.0 \AA\ is greater than unity for the densities and
 temperatures in the core of NGC4636 ISM, while the Fe XVII blend at 17.1 \AA\ is negligible. 
Many scatterings of the 15.0 \AA\ photons flatten its profile comparing to the others, which would cause large uncertainty on Fe abundance.
To correct the optical depth effect at 15.0 \AA, we fit again the 3.5\arcmin\ apeture RGS spectra of those very 
bright galaxies (NGC1399, NGC 4406, NGC 4472, NGC4636, NGC 4649, and NGC 5044), excluding the problematic 15.0 \AA\ line 
(i.e. excluding 14.75 -- 15.4 \AA~) as well as freeing the redshift parameter to account any residual line shift. 
It turns out that only NGC 4636 shows large changes in both parameter values 
(18\% increase in Fe, and 33\% increase in O) and in reduced $\chi_r^2$ (decrease by 23\% ).
Other galaxies only show at most 4\% variances in Fe, O, and reduced $\chi_r^2$. So 
optical depth effect at 15.0 \AA\ line is negligible for all the galaxies except for NGC 4636. After corrected 
this effect for NGC 4636, we still have slightly lower abundances of Fe ($1.07^{+0.11}_{-0.02}$) and O ($0.57^{+0.03}_{-0.07}$) from 1T fit, but have consistent abundances of Fe ($1.42^{+0.16}_{-0.17}$) and O ($0.58^{+0.05}_{-0.10}$) from the 2T fit, comparing to Fe (1.29 $\pm$ 0.06) and O (0.63 $\pm$ 0.06) from \citet{xu02}. These abundances are slightly higher than what {\it XMM\/} EPIC data and {\it ASCA\/} data reveal. \citet{buote99} obtained abundances of Fe and O with {\it ASCA\/} of $1.08^{+0.41}_{-0.24}$ and $0.49^{+0.29}_{-0.18}$, respectively, which are consistant with our 2T fit for {\it XMM\/} EPIC data at 4\arcmin\ diameter region, $1.06\pm0.03$ for Fe and $0.51\pm0.03$ for O. The reason that these abundances from CCD spectral analysis are systematically lower (25\% lower for Fe and 12\% lower for O) than the {\it XMM\/} RGS results is not due to the uncorrected optical depth effect for Fe XVII line at 15 \AA, since we refitted the {\it XMM\/} EPIC 4\arcmin\ diameter region excluding this line (i.e. excluding 0.79 -- 0.88 kev), and found only 4\% increase in Fe and 2\% increase in O. The reason might be that the RGS and EPIC cameras cover different regions of this galaxy, which shows complex structure and events that have been described by others \citep{jones2002, osul05}.

\subsection{Comparison with other studies of the metal abundances in early-type galaxies}
We show below the detailed comparison with other studies of the metal abundances in early-type galaxies for individual galaxy. Note that our fitted temperatures agree well with the published results, and abundances from literatures are scaled to \citet{grev98}.

NGC 720 -- Our 4\arcmin\ diameter aperture EPIC results for NGC 720 (Fe($1.10^{+0.11}_{-0.26}$), O($0.45^{+0.22}_{-0.14}$), Mg($0.76^{+0.73}_{-0.32}$), and Si($0.97^{+0.93}_{-0.45}$)) are consistent with {\it Chandra\/} results reported by \citet{hump06} (Fe($0.71^{+0.40}_{-0.21}$), O($0.16 \pm 0.16$), and Mg($0.90 \pm 0.37$)), and are also consistent with the 180 ks Suzaku observation by \citet{tawara08} within 90\% error range, who obtained abundances of Fe($0.73^{+0.11}_{-0.08}$), O($0.47^{+0.11}_{-0.10}$), Mg($0.50^{+0.11}_{-0.09}$), and Si($0.54^{+0.18}_{-0.15}$).

NGC 1399 -- This is an important galaxy with supersolar abundances reported by previous observations. \citet{buote99} found an Fe abundance of $2.40^{+1.10}_{-0.90}$ within a 10\arcmin\ diameter aperture by {\it ASCA\/}. Later, he reported an {\it XMM\/} result for this galaxy that the Fe abundance in the central region (about 10\arcmin\ in diameter) is 1.54--2.04 \citep{buote02}. Our net 120 ks {\it XMM\/} observation also shows a consistent but more constrained Fe abundance, $1.32\pm0.05$, within 4\arcmin\ diameter region. This Fe abundance is higher than what \citet{hump06} obtained with {\it Chandra\/} data, $1.06\pm0.09$. This is because \citet{hump06} obtained an emission-weighted average Fe abundance up to 20\arcmin\ in diameter, which tends to be lower because the Fe abundance decreases to subsolar at large radius (20\arcmin\ in diameter) of NGC 1399 \citep{hump06}.

NGC 3923 -- \citet{hump06} reported a poorly constrained abundance of Fe for 1.03($>0.24$) due to very few counts of photons. We obtained a lower but more constrained Fe abundance ($0.72^{+0.25}_{-0.19}$). Since we have many more counts of photons (total counts of $3.3*10^4$) for this galaxy and obtained a good fit ($\chi_r^2=1.024$), we are confident with our result.

NGC 4406 -- {\it ASCA\/} observation shows an Fe abundance of $0.77^{+0.33}_{-0.27}$ \citep{buote98}, while {\it Chandra\/} observation shows Fe and O abundances of $1.10^{+0.06}_{-0.47}$ and $0.40^{+0.13}_{-0.24}$, respectively \citep{athey07}. Our results are consistent with these two within error range, which are Fe of $0.87^{+0.02}_{-0.04}$ and O of $0.51^{+0.05}_{-0.05}$.

NGC 4472 -- \citet{buote99} reported an {\it ASCA\/} result for this galaxy to be $2.96^{+3.55}_{-1.44}$ for Fe abundance, while \citet{hump06} reported an emission-weighted average abundances for Fe and O of $1.25^{+1.52}_{-0.36}$ and $0.48^{+0.18}_{-0.18}$, respectively. Our {\it XMM\/} result gives consistent but more constrained Fe and O abundances of $1.75^{+0.07}_{-0.06}$ and $0.77^{+0.06}_{-0.05}$, respectively. 

NGC4552 -- \citet{hump06} measured abundances of Fe and O from {\it Chandra\/} data of $0.63^{+0.24}_{-0.09}$ and $0.08\pm0.06$, while \citet{athey07} reported abundances also measured from {\it Chandra\/} data up to 2.5\arcmin\ diameter region to be $0.73^{+0.07}_{-0.67}$ for Fe and $0.29^{+0.04}_{-0.18}$ for O. We obtained better constrained abundances of Fe and O of $0.49\pm0.08$ and $0.25^{+0.07}_{-0.05}$, respectively, which are consistent with \citet{athey07}'s Fe and O abundances within 90\% error range, but is 68\% higher in O abundance than \citet{hump06}'s result. We attribute this discrepency to be not having enough counts in {\it Chandra\/}'s spectra to constrain the O abundance.

Extra comment for NGC4552: This system is known to be suffering fairly strong ram-pressure stripping and the extent of its gas is quite well-defined from the Chandra images \citep{machacek06}. The 1\arcmin\ diameter region is quite a good match to the gas, but the 4\arcmin\ aperture might contain a large fraction of cluster and mixed cluster/galaxy gas. In our two-phase model fits in Table 3, however, the abundances are quite consistent between 1\arcmin\ and 4\arcmin\ regions. Therefore, the cluster emission will not significantly affect the measurement of the galaxy hot gas abundances in NGC4552 even for 4\arcmin\ apeture. The reason may be that even for the 4\arcmin\ region, the local hot gas in NGC4552 is still the dominant thermal component comparing to other thermal component (like the cluster emission), which contributes 63\% of the total thermal luminosity.

NGC 4636 -- See detailed discussion in Section~\ref{sec:RGScompare}.  

NGC 4649 -- See detailed discussion in Section~\ref{sec:cross-compare}.

NGC 5044 -- \citet{tamu03} obtained abundances of Fe and O within the central 10-20 kpc region (2\arcmin\ width) of NGC 5044 from {\it XMM-Newton\/} RGS data. Their abundances for Fe and O are $0.81\pm0.07$ and $0.31\pm0.13$, respectively, which is consistent in O with but higher in Fe than our RGS 1-T model fitting results for 1\arcmin\ region, Fe($0.65^{+0.09}_{-0.07}$), O($0.42^{+0.10}_{-0.08}$), but are consistent with our RGS 2-T model fitting results of Fe($0.77^{+0.14}_{-0.12}$) and O($0.48^{+0.12}_{-0.11}$). \citet{athey07} used {\it Chandra\/} for annualar analysis for this galaxy, and found abandances of Fe, O, Mg, and Si within 4\arcmin\ diameter in the ranges of 0.92--1.53, 0.18--0.96, 1.01--1.77,  and 0.95--1.93, respectively. These abundances gradients are also conformed by our study with two aperture analysis of EPIC data. We find abundances gradients of Fe, O, Mg, and Si betweem smaller and larger apertures in the ranges of 0.86--1.15, 0.41--0.49, 0.84--1.14, and 0.86--1.01, respectively. Our average Fe abundance ($1.15\pm0.05$) within 4\arcmin\ diameter is also consistent with the result by \citet{buote03}, who reported the {\it XMM\/} observation of emission-weighted average abundance of Fe within 10\arcmin\ diameter to be $1.09\pm0.04$.

IC 1459 -- This is the only galaxy in our sample that shows different spectral charactistics due to a strong central point source \citep{Fabbiano03}. Its power-law component is extremely large compared to the thermal component, making it hard to constrain the abundances in the thermal emission. We fit its spectra excluding the central 13\arcsec\ region in radius, which is mainly a power-law component with a photon index $\Gamma$ of 1.88 and an excessive absorption $n_H$ of $2.9\times10^{21} cm^{-2}$\citep{Fabbiano03}. We then fit the remainder (mainly thermal emission) with a 1-T thermal component plus a power-law (phont index $\Gamma$ fixed to 1.56) with only Fe allowed to vary (other elements are tied to Fe). This gives a better constrain on the Fe abundance of $0.39^{+0.14}_{-0.09}$, which is consistent with the {\it Chandra\/} result of $0.74^{+1.61}_{-0.44}$ \citep{athey07} and the {\it ASCA\/} result of $0.21^{+0.99}_{-0.13}$ \citep{buote98}, within their error ranges. Longer observation is needed to constrain other element abundances. To test if the central point source contributes any photons outside the excluded 13 \arcsec\ region, we refitted IC1459 EPIC 1\arcmin\ region excluding the central 13\arcsec\ core with photon index free to fit. The fitted photon index is $1.60_{-0.05}^{+0.06}$, only increased by 2.6\% from 1.56, and the temperature and Fe abundance change not much, by 0.5\% and 3.7\%, respectively. For 4\arcmin\ region, the results are essentially not changed. So we conclude that the central point source in IC1459 doesn't contribute many photons outside the excluded 13" region.

\subsection{Discussion on the Assumptions of the Spectral Fitting Method}
In our spectral fitting method, we made two simplifying assumptions in the choice of spectral models - freezing the slope of the power law component and linking the abundances of the two-temperature models. We did the following tests to justify these assumptions.

{\textbf{Effect on the slope of the power law component:}} To check the effect on photon index, we test two cases (with 2T model plus a power law), NGC5044 and NGC3923, according to their different fractions of the total luminosity for the power-law component. During the fitting, the photon index is free to fit.

For NGC5044, the power-law component only contributes 3.4\% to the total luminosity in the energy band of 0.33 kev to 2.48 kev in its central 1\arcmin\ diameter region. The fitted parameters don't change much. The result shows a best-fit photon index of $1.80^{+0.27}_{-0.28}$, which is consistent with 1.56 within its 90\% error range. The two temperatures vary by only 0.3\%, while the abundances vary at most 5.8\%, which are within 90\% confidence error range of previous result when fixed photon index to 1.56. 

For NGC3923, the power-law component contributes 24.6\% to the total luminosity in the energy band of 0.33 kev to 2.48 kev in its central 4\arcmin\ diameter region. The fitted photon index is $1.73_{-0.20}^{+0.09}$, which is consistent with 1.56 within its 90\% error range. There are no significant changes in reduced chi-sqaure value as well as the two temperature values, only by 0.4\%. Abundance values, however, show large increase by at least 50\%. Fe changes most, increased by 70.8\%, from $0.72_{-0.19}^{+0.25}$ to $1.23_{-0.44}^{+1.07}$, but is still consistent within their error range.

So we conclude that the value of photon index doesn't significantly affect the abundance values for hot gas dominent galaxies, and freezing photon index to 1.56 can better constrain the abundance values. But it is the main error source for galaxies where power-law component is comparable to the hot gas component. In our sample, most sources only have a few percent of the total luminosity for the power-law component, and thus are not greatly affected by the value of photon index. Only four galaxies (NGC720, NGC3923, NGC4552, and IC1459) have power-law component with luminosity greater than 15\% of the total luminosity, thus their abundances are more greatly affected by the value of photon index. Longer observations for these sources are needed to better constrain the abundances.

{\textbf{Effect on linking the abundances of the 2T models:}} There are five galaxies (NGC1399, NGC4406, NGC4472, NGC4649, and NGC5044) in our sample which show the possible signs of overlying cluster emission. This can be seen from the 2T model fitting temperatures (see Table 3), where the galaxy hot gas temperatures are around 0.63-0.81 kev, while the cluster emission temperatures are around 0.94-1.72 kev.

 We chose NGC1399 4\arcmin\ region to test the effect of linking two abundance sets, because it has the most photon counts and a comparable cluster emission component to the local galaxy hot gas (about 46\% of the total thermal luminosity). During the fit, we unlinked the two abundance sets. Each abundance set then has free parameters of Fe, O, Mg, Si, S, Ne, and Ni. Other elements are tied to Fe. The fitting results show no change for the reduced chi-sqare value, but the 90\% confidence range for each abundance parameter becomes larger. The best-fit parameters for the lower temperature component are T:$0.811^{+0.004}_{-0.004}$, Fe:$1.89^{+0.67}_{-0.57}$, O:$1.29^{+0.43}_{-0.69}$, while values for the higher temperature componet are T:$1.445^{+0.05}_{-0.05}$, Fe:$1.20^{+0.47}_{-0.13}$, and O:$0.10_{-0.10}^{+0.68}$. These two sets of abundances do not show much difference within their 90\% error bars. They are also consistent with the best-fit parameters when linking the abundances of the 2T models.

So in our sample, even for the region that has the most contamination from the cluster emission, linking the two abundance sets would not significantly affect the hot gas abundances, and it can better constrain the abundances by using fewer free parameters. Direct measurement of abundances for different thermal components are not good for low quality spectra, and deprojected method should be used. 

\subsection{Discussion on the Effect of RGS Background}
 For the 1\arcmin\ RGS region, the local galaxy hot gas emission is always the dominant component. Thus it makes little difference, whether using an RGS background template or using one that contains the underlying cluster emission but free of lcoal galaxy hot gas emission. 

For the 3.5\arcmin\ RGS region, however, the background could have some effect, particularly for those with the brightest surrounding group/cluster emission. In our sample, fainter systems like NGC4552, NGC720, NGC3923, and IC1459 have too poor RGS spectra to constrain their abundances, even for the 3.5\arcmin\ region. For brighter systems, however, it is very hard to get a local background that contains only the underlying cluster emission but free of galaxy hot gas emission. This is because their extended hot gas emission regions are usually larger than 5\arcmin\ in diameter (checked with XMM MOS images), but the maximum field of view of RGS in the cross-dispersion direction is only 5\arcmin. The only way to obtain this local background is using seperate observation pointing off the target and on the surrounding cluster emission, which is not available for all the galaxies in our sample. Thus using local contaminated background will overestimate the background, while using the RGS background template will underestimate the underlying cluster emission. This is the main error source when determine the abundances for the RGS 3.5' region, especially when the cluster hot gas component is comparitable to the galaxy hot gas component, which can be seen by the luminosity ratio of the two thermal components obtained from the EPIC fits.

\section{Discussion and Conclusions}

There have been several previous determinations of metal abundances in
early-type galaxies, some of which have obtained similar results \citep{buote99,hump06,athey07}. 
One important difference in this work is that, for a sample of 10 galaxies, 
we include the RGS data, in which
one can clearly see the strong individual lines from Fe, O, and other species,
whereas they are thoroughly blended in the {\it XMM\/} EPIC and in the {\it Chandra\/} ACIS data.
The identification of these ions not only gives us greater confidence in the
abundance determinations, but they often provide very strict constraints on the
maximum allowable abundances.

Of the models considered, the one yielding the lowest $\chi^2$ has a power-law
to represent unresolved X-ray binaries, two thermal components with one set of
variable abundances for several individual elements (Fe, O, Mg, and Si), and 
Galactic absorption.  This model led to formally acceptable or good fits for 
most of the galaxies except for NGC 4636 and IC 1459.  In IC 1459, there is a 
strong central point source that compromised spectral fits to the fainter emission
from the hot gas.  In NGC 4636, a two-temperature model is inadequate, possibly
due to the structure and events that have been described by others \citep{jones2002, osul05}.
Excluding NGC 4636 and IC 1459, the remaining eight galaxies have median
abundances (25\% and 75\% quartile points are given in parenthesis for Fe and O) 
relative to Solar values \citep{grev98} of 0.86 (0.64-1.70), 0.44 (0.29-0.60), 0.81,
and 0.79 for Fe, O, Mg, and Si.  The abundances for Mg and Si are not distinuishable
from that of Fe.  Three of the galaxies have mildly supersolar Fe abundances, at
about 1.5 of the Solar value, and we note that the three all lie at the center
of their group or cluster (NGC 1399 in the Fornax cluster; NGC 4472 is the dominant
galaxy in the southern part of the Virgo cluster; NGC 4649 is the brightest member
of a group on the outskirts of the Virgo cluster).  While this may be a trend,
we note that the galaxy NGC 5044 is also the brightest member of its group yet it has 
abundances that are close to the median values.
The galaxy that has the lowest Fe abundance is NGC 4552, the optically least luminous galaxy of
this group (and relatively lower in X-ray luminosity).  As we will discuss in
a future paper, relatively low abundances are found frequently in galaxies with
lower X-ray and optical luminosities, as \citet{osul04} has found.

One of the issues that we examined was the difference between abundances 
when a second temperature component is added, sometimes referred to as 
the Fe-bias \citep{trinc94, buote2000}.  The need for a second temperature
component is probably due to a temperature gradient within the galaxy.
Upon adding the second thermal component, the increase in Fe was about
30\% (quartile values of 7-38\%) and for O it was 4\% (quartile values of 
-3\%$\sim$16\%).  There are two galaxies in which there was essentially no change
in the abundances, NGC 4649 and NGC 5044.  From radial temperature profiles,
NGC 4649 is known to be nearly isothermal \citep{athey07, rand06},
which explains why adding a second temperature component was of little consequence.  
For NGC 5044, there was always a dominant 
thermal component and the luminosity from an additional thermal component was
relatively small in the best-fit solutions.

Because NGC 4649 can be adequately fit with a single-temperature component and the
temperatures do not change with an additional thermal component, we used this 
source to compare results between spectral fits for the three {\it XMM\/} instruments 
(EPIC MOS, EPIC PN, and the RGS) and for the {\it Chandra\/} ACIS-S.  The derived 
temperatures were the same to about 1\%, and there was excellent agreement
between the three {\it XMM-Newton\/} instruments for the seven fitted abundances.
The abundances derived from the {\it Chandra\/} ACIS-S were also in good agreement
for Fe, Mg, Si, and Ni, but gave lower values than the {\it XMM-Newton\/} data for
O, Ne, and S.  Overall, Fe, Mg, and Si had similar abundances that were 1.5
times the Solar value, S is consistent with that value, and Ni may be supersolar,
although the constraints are poor.  Oxygen, the lightest metal in the group, 
has an abundance of about 0.6 Solar and Ne, the next lightest metal, has an
abundance between O and Fe, at about 0.7 Solar.

\begin{figure}
\epsscale{1.0}
\plotone{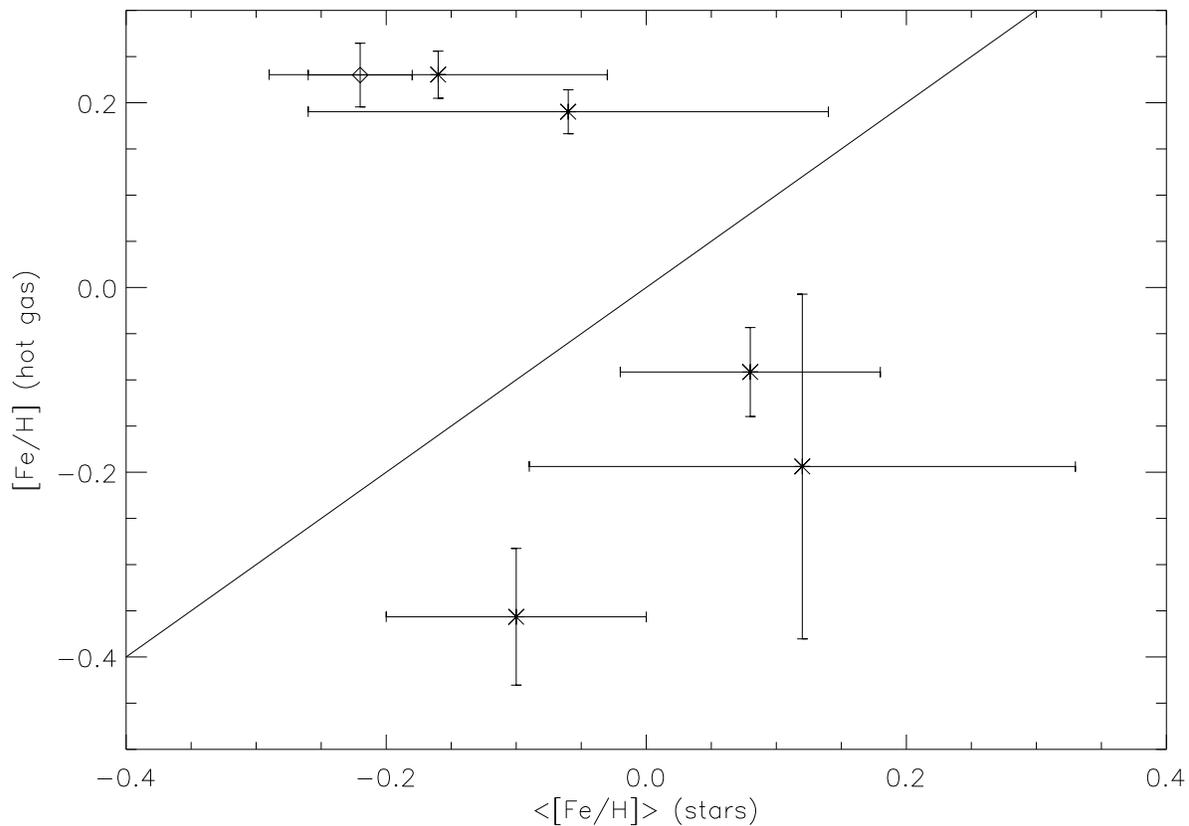}
\caption[x-ray VS optical]%
        {Comparison of Fe abundances from hot gas (our results) and from optical studies of stars \citep{hump06,trager00}. Symbol X: mean stellar Fe abundances from \citet{hump06}; symbol diamond: mean stellar Fe abundance from \citet{trager00}. The solid line represents where the hot gas and the stars would have the same Fe abundance (it is not a fit to the data).}
\label{fig:optFe}
\end{figure}

Metals in hot gas of early type galaxies are thought to come from stellar mass loss, along with supernovae. Since supernovae mainly contribute heavy elements to the hot gas, we should expect supersolar metalicity in hot gas. In Fig.~\ref{fig:optFe}, we compared Fe abundances found in hot gas and those from optical studies (only six galaxies in our sample have optical Fe abundances reported in literature by \citet{hump06} and \citet{trager00}). Only three galaxies (NGC 1399, NGC 4472, and NGC 4649) show modest supersolar Fe abundances in hot gas, while the other three (NGC 720, NGC 3923, and NGC 4552) are slightly subsolar. Supernovae were expected to enhance the gas phase abundances considerably, which is not observed in our sample. 
 
 Element abundance ratios can constrain the relative contributions of SN Ia or SN II to the metallicity. In our sample, the median abundance ratios are 0.51, 0.94, and 0.92 for O/Fe, Mg/Fe, and Si/Fe, respectively. These give us the fraction of contribution from Type Ia supernovae of 64\%-85\% when using linear combination of SN Ia yeilds from W7 model and SN II yields from \citet{nomoto97}, which indicates that most of the metal enrichment is from Type Ia supernovae.

One important result is the low O/Mg ratio, which we derived a value of 0.54, which is also confirmed to be low by \citet{athey07} and \citet{hump06}. Theoretically, One should expect similar abundance of O to Mg, since SN Ia produces similar amount of O and Mg while SN II produce slightly more O than Mg. The chemical studies of Galactic bulge stars, which are also old star population similar to stars in elliptical galaxies but are mainly enriched by Type II supernovae, also show a similar low O/Mg ratio at high Fe abundance \citep{Fulbright04,Minniti07}. This low O abundance in the hot gas as well as in bulge stars might be the evidence of an overestimate of O yield by SN II models, which do not consider significant mass loss at the late stage of massive progenitor stars. Since strong stellar wind from massive progenitor stars might reduce or even completely blow away their He-rich envelopes while still preserve the C-O layers, thus after Type II supernove explosions, only $\alpha$ elements (like Mg) by Carbon burning are synthesized while elements like O through hydrostatic He burning are greatly reduced \citep{Fulbright04}. The observed Mg-rich SN remnent of N49B without accompanying Ne or O can support this argument \citep{park03}.     

\acknowledgements
We would like to thank Renato Dupke for his comments and suggestions.
We gratefully acknowledge financial support for this research, which was provided by NASA.

\clearpage



\begin{deluxetable}{l c c c c c c c c c c c c c c c}
\tabletypesize{\tiny}
\tablecaption{Properties and Observational Information of Surveyed Galaxies}
\tablehead{
\multirow{2}{*}{Galaxy}& \multirow{2}{*}{Type\tablenotemark{a}} & \colhead{$B_T^0$\tablenotemark{b}} &\colhead{ D \tablenotemark{c}}  & \colhead{$N_H$ \tablenotemark{d}}& \colhead{$r_e$\tablenotemark{e}} & \multirow{2}{*}{ObsID}\tablenotemark{f} & \multicolumn{6}{c}{Net exposure time (ks)\tablenotemark{g}}\\
   \colhead{}& \colhead{} &      \colhead{ (mag)}& \colhead{(Mpc)} & \colhead{($10^{20} cm^{-2}$)} & \colhead{(arcsec)}&\colhead{}  & \colhead{MOS1} & \colhead{MOS2} & \colhead{PN} & \colhead{RGS1} & \colhead{RGS2} & \colhead{ACIS-S}\\
}
\startdata

  NGC720 & E5 & 11.13 & 27.67 & 1.55 & 39.87 & 0112300101 & 29.6 & 30.0 & 19.9 & 46.3 & 45.3  & - \\ 
  NGC1399 & cD;E1pec & 10.44 & 19.95 & 1.31 & 42.55 & 0400620101 & 121.5 &120.5  & 73.9 & 118.2 & 118.1 & -   \\ 
  NGC1399     &  -  &    -  &    -  &   -   &   -    &   319 $^h$  &- & - & - & - & - & 55.9\\
  NGC3923 & E4-5 & 10.62 &22.91 & 6.29 &53.35 & 0027340101 & 38.8 & 38.7 & 29.8 & 43.9 & 42.6  & - \\ 
  NGC4406 & S0(3)/E3 & 9.74 & 17.14 & 2.58 & 89.64 & 0108260201 & 77.9 & 78.9 & 47.8 & 83.5 & 81.1 & -  \\ 
  NGC4472 & E2/S0 & 9.33 & 16.29& 1.65 & 104.40 & 0200130101 & 82.5 & 82.7 & 72.8 & 101.5 & 101.5  & - \\ 
  NGC4472       &  -  &    -  &   -   &  -    &    -   &   321 $^h$   &- & - & - & - & - & 32.5 \\
  NGC4552 & E & 10.57 & 15.35 & 2.56 & 48.89 & 0141570101 & 27.8 & 31.2 & 18.5 & 42.9 & 42.8  & - \\ 
  NGC4636 & E/S0\_1 & 10.43 & 14.66 & 1.83 & 100.08 & 0111190701 & 59.2 & 59.3 & 51.1 & 62.9 & 61.4 & -  \\ 
  NGC4649 & E2 & 9.70 & 16.83 & 2.13 & 73.73& 0021540201 & 50.6 & 50.6 & 42.2 & 53.1 & 51.6  & - \\ 
  NGC4649     &  -  &    -  &    -  &     - &    -   &      785$^h$  &- & - & - & - & - &  22.9 \\
  NGC5044 & E0 & 11.67 & 31.19 & 5.03 & 82.23 & 0037950101 & 22.6 & 22.7 & 17.0 & 23.6 & 22.8  & - \\ 
  IC1459 & E3 & 10.83 &29.24  & 1.19 & 38.61 & 0135980201 & 29.3 & 29.3 & 25.2 & 31.7 & 30.8 & -  \\ 
 
\enddata

\tablenotetext{a}{The galaxy type was take from NED.}
\tablenotetext{b}{Total B band magnitude from RC3 (\citet{devauc91}).}
\tablenotetext{c}{Distances in Mpc, measured by surface brightness fluctuation method (\citet{tonry01}).}
\tablenotetext{d}{The Galactic H I column density, taken from the dust map by \citet{dick90}.}
\tablenotetext{e}{Effective, blue-half light radius in arcsecond derived from RC3 (\citet{devauc91}).}
\tablenotetext{f}{{\it XMM-Newton\/} observation ID. $^h$:{\it Chandra\/} ObsID.}
\tablenotetext{g}{Net exposure time after filtering background flares.}
\label{tab:obs}
\end{deluxetable}



\begin{deluxetable}{l c c c c c c c c c c c c c c}
\tabletypesize{\tiny}
\setlength{\tabcolsep}{2pt}
\tablecaption{One Phase Model Fitting for {\it Chandra} ACIS-S, {\it XMM\/} EPIC, and RGS spectra.}
\tablehead{
 \colhead{Galaxy}&  \colhead{Camera/Size\tablenotemark{a}} & \colhead{$T$(keV)\tablenotemark{b}} &\colhead{ Fe \tablenotemark{c}}  & \colhead{O \tablenotemark{c}}& \colhead{Mg\tablenotemark{c}} & \colhead{Si\tablenotemark{c}}& \colhead{Ne\tablenotemark{c}}& \colhead{S\tablenotemark{c}}& \colhead{Ni\tablenotemark{c}} & \colhead{dof\tablenotemark{d}}& \colhead{$\chi^2_r$\tablenotemark{e}}& \colhead{Power\tablenotemark{f}}& \colhead{Vapec\tablenotemark{g}}&  \colhead{Vapec/Power\tablenotemark{h}}\\
}
\startdata

NGC720   &EPIC 1\arcmin & $0.560_{-0.008}^{+0.015}$ &   $0.51_{-0.05}^{+0.06}$ & $0.30_{-0.13}^{+0.17}$ & $0.40_{-0.16}^{+0.19}$ & $0.35_{-0.25}^{+0.30}$ &$0.54 _{-0.29}^{+0.30}$& - & $1.26_{-0.98}^{+1.02}$ & 198 & 1.102 & 0.54 & 1.73 &  3.22 \\ 
  &EPIC 4\arcmin& $0.396_{-0.021}^{+0.013}$ &   $1.52_{-0.41}^{+1.27}$ & $0.49_{-0.16}^{+0.54}$ & $1.35_{-0.46}^{+1.54}$ & $2.48_{-1.09}^{+2.52}$ & $2.21_{-0.15}^{+0.15}$ & - & - & 500 & 1.274 & 1.61 & 4.19 &  2.61 \\ 
  &RGS 1\arcmin& $0.369_{-0.057}^{+0.113}$& $1.54_{-1.06}^{+3.46}$ & $0.10_{-0.00}^{+0.20}$ & - & - & - & - & - & 62 & 1.313 & 0.92 & 2.81 & 3.05 \\ 
  &RGS 3.5\arcmin& $0.327_{-0.028}^{+0.033}$& $4.10_{-3.42}^{+0.90}$ & $0.31_{-0.21}^{+0.29}$ & - & - & - & - & - & 143 & 1.303 & 2.06 & 5.37 & 2.61 \\ 

\hline
 NGC1399 & ACIS-S 1\arcmin & $0.809_{-0.003}^{+0.006}$ & $1.33_{-0.09}^{+0.10}$ & $0.77_{-0.12}^{+0.14}$ & $1.49_{-0.19}^{+0.22}$ & $1.58_{-0.18}^{+0.20}$ & $0.37_{-0.34}^{+0.36}$ & $0.76_{-0.28}^{+0.30}$ & $7.29_{-0.87}^{+1.02}$ &  143 & 3.520 & 0.71 &  7.98 &  11.28  \\
& EPIC 1\arcmin& $0.813_{-0.002}^{+0.002}$ &   $1.25_{-0.04}^{+0.04}$ & $0.54_{-0.07}^{+0.07}$ & $1.65_{-0.11}^{+0.11}$ & $1.40_{-0.09}^{+0.09}$ & $1.97_{-0.12}^{+0.13}$ & $1.13_{-0.12}^{+0.11}$ & - & 775 & 2.393 & 0.63 & 7.68 &  12.18 \\ 
 &EPIC 4\arcmin& $0.950_{-0.001}^{+0.002}$ &   $0.65_{-0.02}^{+0.02}$ & $0.50_{-0.03}^{+0.04}$ & $0.89_{-0.04}^{+0.04}$ & $1.06_{-0.04}^{+0.04}$ & $0.22_{-0.04}^{+0.08}$ & $0.89_{-0.05}^{+0.03}$ & $4.17_{-0.11}^{+0.11}$ & 1367 & 2.939 & 3.46 & 19.94 &  5.77 \\
  &RGS 1\arcmin& $0.813_{-0.014}^{+0.011}$ &   $1.32_{-0.37}^{+0.62}$ & $1.20_{-0.36}^{+0.29}$ & - & - & - & - & - & 1011 & 1.190 & 0.86 & 14.12  & 16.46 \\ 
  &RGS 3.5\arcmin& $0.996_{-0.009}^{+0.008}$ &   $0.68_{-0.06}^{+0.10}$ & $0.55_{-0.07}^{+0.10}$ & $0.44_{-0.10}^{+0.26}$ & $0.98_{-0.20}^{+0.30}$ & - & - & - & 1451 & 1.323 & 4.55 & 26.88  & 5.91 \\ 

\hline
NGC3923 &EPIC 1\arcmin & $0.479_{-0.019}^{+0.016}$ &   $0.49_{-0.04}^{+0.09}$ & $0.26_{-0.07}^{+0.08}$ & $0.45_{-0.10}^{+0.14}$ & $0.63_{-0.19}^{+0.26}$ & $0.27_{-0.14}^{+0.14}$ & - & $3.11_{-0.78}^{+0.77}$ & 315 & 1.346 & 0.61 & 3.48 &  5.68 \\ 
  &EPIC 4\arcmin& $0.387_{-0.014}^{+0.016}$ &   $0.63_{-0.12}^{+0.17}$ & $0.21_{-0.05}^{+0.06}$ & $0.60_{-0.16}^{+0.22}$ & $0.96_{-0.35}^{+0.48}$ & $0.58_{-0.10}^{+0.10}$ & - & $2.73_{-1.08}^{+1.07}$ & 588 & 1.168 & 1.86 & 5.20 &  2.80 \\ 
  &RGS 1\arcmin& $0.446_{-0.077}^{+0.050}$& $5.00_{-3.89}^{+0.00}$ & $2.06_{-1.06}^{+1.33}$ & - & - & - & - & - & 91 & 1.421 & 0.64 & 3.76 & 5.86 \\ 
  &RGS 3.5\arcmin& $0.432_{-0.067}^{+0.057}$& $5.00_{-3.33}^{+0.00}$ & $3.01_{-1.65}^{+1.99}$ & - & - & - & - & - & 184 & 1.476 & 1.44 & 4.66 & 3.24 \\ 

\hline
 NGC4406 &EPIC 1\arcmin& $0.647_{-0.007}^{+0.007}$ &   $0.64_{-0.04}^{+0.04}$ & $0.40_{-0.09}^{+0.10}$ & $0.65_{-0.10}^{+0.11}$ & $0.69_{-0.12}^{+0.13}$  & $0.83_{-0.20}^{+0.19}$ & - & $1.79_{-0.52}^{+0.46}$ & 462 & 1.273 & 0.01 & 0.08 &  8.26 \\ 
 &EPIC 4\arcmin& $0.716_{-0.003}^{+0.003}$ &   $0.65_{-0.03}^{+0.03}$ & $0.49_{-0.05}^{+0.05}$ & $0.65_{-0.05}^{+0.05}$ & $0.62_{-0.05}^{+0.05}$ & $0.82_{-0.10}^{+0.10}$ & $0.58_{-0.10}^{+0.09}$ & $2.33_{-0.22}^{+0.24}$  & 919 & 1.409 & 0.04 & 0.44 &  12.07 \\ 
  &RGS 1\arcmin& $0.626_{-0.018}^{+0.017}$ &   $0.34_{-0.04}^{+0.06}$ & $0.32_{-0.08}^{+0.11}$ & $0.15_{-0.05}^{+0.20}$ & - & - & - & - & 304 & 1.294 & 0.03 & 0.29  & 8.32 \\ 
  &RGS 3.5\arcmin& $0.623_{-0.010}^{+0.011}$ &   $0.29_{-0.02}^{+0.03}$ & $0.25_{-0.04}^{+0.05}$ & $0.16_{-0.06}^{+0.11}$ & - & - & - & - & 774 & 1.230 & 0.06 & 0.80  & 13.29 \\ 

\hline
 NGC4472 &ACIS-S 1\arcmin & $0.724_{ -0.008}^{ +0.015}$ & $1.58 _{-0.18}^{ +0.15 }$&$ 0.71_{ -0.23}^{ +0.29}$ & $1.99_{ -0.34}^{ +0.45}$ &$ 1.87_{ -0.32}^{ +0.41}$  &$2.15_{ -0.86}^{ +1.04}$ &$ 1.00_{ -0.49}^{ +0.58}$  &$3.35_{ -1.59}^{ +1.83}$ & 123  & 1.595  & 0.29   & 4.15  &  14.11\\
&EPIC 1\arcmin& $0.748_{-0.003}^{+0.003}$ &   $1.11_{-0.03}^{+0.06}$ & $0.54_{-0.06}^{+0.07}$ & $1.17_{-0.08}^{+0.09}$ & $1.19_{-0.08}^{+0.09}$  & $0.83_{-0.14}^{+0.15}$ & $0.99_{-0.13}^{+0.13}$ & $3.93_{-0.28}^{+0.32}$ & 720 & 1.771 & 0.40 & 4.36 &  11.01 \\ 
 &EPIC 4\arcmin& $0.813_{-0.002}^{+0.002}$ &   $1.05_{-0.03}^{+0.03}$ & $0.51_{-0.04}^{+0.04}$ & $1.55_{-0.06}^{+0.07}$ & $1.32_{-0.05}^{+0.06}$ & $2.81_{-0.08}^{+0.09}$ & $1.18_{-0.09}^{+0.08}$ & - & 1162 & 5.235 & 1.38 & 10.50 &  7.59 \\
  &RGS 1\arcmin& $0.752_{-0.010}^{+0.010}$ &   $0.71_{-0.11}^{+0.17}$ & $0.59_{-0.05}^{+0.14}$ & $0.53_{-0.18}^{+0.31}$ & - & - & - & - & 536 & 1.478 & 0.56 & 8.37  & 15.01 \\ 
 &RGS 3.5\arcmin& $0.812_{-0.006}^{+0.003}$ &   $1.04_{-0.15}^{+0.29}$ & $0.91_{-0.14}^{+0.24}$ & $0.81_{-0.27}^{+0.48}$ & -& - & - & - & 1015 & 1.231 & 1.50 & 13.56  & 9.06 \\ 

\hline
NGC4552 &EPIC 1\arcmin& $0.608_{-0.010}^{+0.010}$ &   $0.42_{-0.05}^{+0.06}$ & $0.29_{-0.08}^{+0.09}$ & $0.46_{-0.10}^{+0.11}$ & $0.44_{-0.12}^{+0.13}$  & - & - & - & 414 & 1.063 & 0.07 & 0.18 &  2.64 \\ 
&EPIC 4\arcmin& $0.386_{-0.011}^{+0.012}$ &   $0.66_{-0.12}^{+0.20}$ & $0.22_{-0.06}^{+0.08}$ & $0.96_{-0.23}^{+0.33}$ & $1.45_{-0.53}^{+0.72}$ & $1.16_{-0.07}^{+0.07}$ & - & - & 628 & 1.295 & 0.13 & 0.22 &  1.69 \\ 
  &RGS 1\arcmin& $0.557_{-0.054}^{+0.032}$& $4.32_{-3.33}^{+0.68}$ & $4.61_{-4.44}^{+0.39}$ & - & - & - & - & - & 89 & 1.640 & 0.07 & 0.18 & 2.59 \\ 
  &RGS 3.5\arcmin& $0.376_{-0.040}^{+0.049}$& $5.00_{-3.24}^{+0.00}$ & $2.59_{-0.96}^{+1.59}$ & - & - & - & - & - & 190 & 1.469 & 0.15 & 0.25 & 1.67 \\ 

\hline
  NGC4636 &EPIC 1\arcmin& $0.535_{-0.006}^{+0.006}$ &   $0.62_{-0.02}^{+0.01}$ & $0.41_{-0.03}^{+0.04}$ & $0.40_{-0.04}^{+0.04}$ & $0.61_{-0.06}^{+0.06}$  & $0.24_{-0.06}^{+0.06}$ & - & $2.35_{-0.25}^{+0.31}$ & 506 & 2.061 & 0.14 & 4.83 &  34.70 \\ 
  &EPIC 4\arcmin& $0.631_{-0.001}^{+0.002}$ &   $0.92_{-0.03}^{+0.03}$ & $0.58_{-0.03}^{+0.03}$ & $0.69_{-0.03}^{+0.03}$ & $0.91_{-0.04}^{+0.04}$ & $0.74_{-0.06}^{+0.06}$ & - & $1.06_{-0.20}^{+0.19}$ & 838 & 2.167 & 0.53 & 14.89 &  27.92 \\  
  &RGS 1\arcmin& $0.622_{-0.008}^{+0.009}$ &   $0.83_{-0.10}^{+0.10}$ & $0.38_{-0.07}^{+0.08}$ & $0.45_{-0.16}^{+0.18}$ & - & - & - & - & 438 & 2.020 & 0.20 & 6.93  & 34.92 \\ 
  &RGS 3.5\arcmin& $0.639_{-0.006}^{+0.006}$ &   $0.91_{-0.07}^{+0.10}$ & $0.43_{-0.06}^{+0.07}$ & $0.48_{-0.13}^{+0.16}$ & - & - & - & - & 893 & 1.700 & 0.52 & 14.57  & 27.91 \\ 
  &RGS corrected 1\arcmin\tablenotemark{i}& $0.589_{-0.008}^{+0.009}$ &   $1.00_{-0.05}^{+0.12}$ & $0.53_{-0.08}^{+0.04}$ & $0.65_{-0.21}^{+0.14}$ &- & - & - & - & 389 & 1.342 & 0.20 & 7.19  & 36.22 \\ 
  &RGS corrected 3.5\arcmin\tablenotemark{i}& $0.625_{-0.006}^{+0.007}$ &   $1.07_{-0.02}^{+0.11}$ & $0.57_{-0.07}^{+0.03}$ & $0.63_{-0.17}^{+0.12}$ &- & - & - & - & 810 & 1.307 & 0.52 & 14.85  & 28.48 \\ 

\hline
 NGC4649 &ACIS-S 1\arcmin & $ 0.763_{ -0.017}^{ +0.014}$& $ 1.39_{ -0.19}^{ +0.21}$& $ 0.13_{ -0.13}^{ +0.17}$& $ 1.46_{ -0.28}^{ +0.29}$& $ 1.47_{ -0.27}^{ +0.29}$& $ 0.00_{ -0.00 }^{+0.55}$& $ 0.31_{ -0.31 }^{+0.19}$& $ 4.16_{ -1.47}^{ +1.60}$&  112&  1.421 & 0.48&   5.68 &  11.91 \\
&EPIC 1\arcmin& $0.776_{-0.003}^{+0.003}$ &   $1.57_{-0.11}^{+0.12}$ & $0.55_{-0.10}^{+0.11}$ & $1.65_{-0.16}^{+0.18}$ & $1.63_{-0.15}^{+0.17}$  & $0.77_{-0.24}^{+0.12}$ & $1.25_{-0.20}^{+0.20}$ & $4.30_{-0.43}^{+0.45}$ & 630 & 1.408 & 0.58 & 5.66 &  9.73 \\ 
 &EPIC 4\arcmin& $0.784_{-0.002}^{+0.002}$ &   $1.39_{-0.08}^{+0.08}$ & $0.58_{-0.07}^{+0.08}$ & $1.43_{-0.11}^{+0.12}$ & $1.64_{-0.11}^{+0.12}$ & $0.70_{-0.17}^{+0.16} $ & $1.09_{-0.15}^{+0.15}$ & $3.63_{-0.30}^{+0.16}$ & 931 & 1.506 & 1.75 & 9.63 &  5.50 \\ 
  &RGS 1\arcmin& $0.759_{-0.010}^{+0.010}$ &   $1.37_{-0.20}^{+0.55}$ & $0.67_{-0.07}^{+0.33}$ & - & - & - & - & - & 325 & 1.335 & 0.71 & 6.93  & 9.73 \\ 
 &RGS 3.5\arcmin& $0.776_{-0.009}^{+0.014}$ &   $1.53_{-0.14}^{+0.48}$ & $0.96_{-0.21}^{+0.36}$ & $1.01_{-0.42}^{+0.32}$ & - & - & - & - & 552 & 1.258 & 1.66 & 9.41  & 5.67 \\ 

\hline
NGC5044 &EPIC 1\arcmin& $0.751_{-0.005}^{+0.002}$ &   $0.82_{-0.06}^{+0.06}$ & $0.44_{-0.08}^{+0.09}$ & $0.80_{-0.10}^{+0.11}$ & $0.80_{-0.09}^{+0.10}$  & $0.44_{-0.18}^{+0.18}$ & $0.71_{-0.15}^{+0.15}$ & $2.24_{-0.36}^{+0.37}$ & 477 & 1.118 & 1.83 & 49.05 &  26.87 \\ 
 &EPIC 4\arcmin& $0.792_{-0.002}^{+0.002}$ &   $0.72_{-0.02}^{+0.02}$ & $0.37_{-0.04}^{+0.04}$ & $0.94_{-0.05}^{+0.05}$ & $0.89_{-0.04}^{+0.04}$ & $0.56_{-0.08}^{+0.08}$ & - & $3.48_{-0.15}^{+0.15}$ & 832 & 1.963 & 8.80 & 249.84 &  28.39 \\  
 &RGS 1\arcmin& $0.771_{-0.011}^{+0.010}$ &   $0.65_{-0.07}^{+0.09}$ & $0.42_{-0.08}^{+0.10}$ & $0.89_{-0.30}^{+0.33}$ & - & - & - & -  & 295 & 1.262 & 4.60 & 123.97  & 26.95 \\ 
  &RGS 3.5\arcmin& $0.792_{-0.007}^{+0.007}$ &   $0.59_{-0.04}^{+0.05}$ & $0.46_{-0.06}^{+0.07}$ & $0.49_{-0.14}^{+0.14}$ & $0.85_{-0.19}^{+0.19}$ & - & - & - & 668 & 1.210 & 10.25 & 293.81  & 28.65 \\ 

\hline
 IC1459  
  &EPIC 1\arcmin no core\tablenotemark{j}& $0.612_{-0.036}^{+0.035}$ &  $1.62_{-0.25}^{+0.25}$ & - & - & -  & - & - & - & 171  &0.828 &  0.90&     0.31 & 0.35  \\
  &EPIC 4\arcmin no core\tablenotemark{j}& $0.583_{-0.020}^{+0.020}$ &  $0.39_{-0.09}^{+0.14}$ & - & - & - & - & - & - & 462  &1.049 &  1.96&     1.36 & 0.69  \\
  &RGS 1\arcmin& $0.532_{-0.089}^{+0.063}$& $1.18_{-0.84}^{+3.40}$ & $2.32_{-1.79}^{+2.68}$ & - & - & - & - & - & 66 & 1.105 & 2.02 & 1.95 & 0.97 \\
  &RGS 3.5\arcmin& $0.307_{-0.032}^{+0.114}$& $4.96_{-4.66}^{+0.04}$ & $1.84_{-1.64}^{+3.16}$ & - & - & - & - & - & 138 & 0.994 & 2.86 & 2.82 & 0.99 \\ 

\enddata

\tablenotetext{a}{Instruments used and their apertures. EPIC:joint fit for MOS1,MOS2,\& pn data; RGS: joint fit for RGS1 \& RGS2 data. EPIC source size: diameter of 1 \arcmin\ and 4 \arcmin\ ; RGS source size: including 90\% and 99\% of psf, around 1 \arcmin\ and 3.5 \arcmin\ in the cross-dispersion direction. One phase model: {\it wabs(vapec+power)} for EPIC, and {\it wabs(vapec+power)*rgsxsrc} for RGS.}
\tablenotetext{b}{One phase temperature in unit of {\it keV}.}
\tablenotetext{c}{Elemental abundances of Fe, O, Mg, Si, Ne, S, and Ni relative to their Solar values (Grevess et al, 1998). Note: for RGS data, the Mg and Si abundances were allowed to fit when thawing them can improve the fits, otherwise they were tied to Fe.}
\tablenotetext{d}{Degrees of freedom.}
\tablenotetext{e}{Reduced $\chi^2$ of the fit.}
\tablenotetext{f}{{Luminosity} of the power-law component in unit of $10^{40} ergs/s$ and in the range of 0.33 keV to 2.48 keV.}
\tablenotetext{g}{{Luminosity} of the thermal component (vapec) in unit of $10^{40} ergs/s$ and in the range of 0.33 keV to 2.48 keV.}
\tablenotetext{h}{{Luminosity} ratio of the thermal component to the power-law one.}
\tablenotetext{i}{We excluded the Fe XVII line at 15 \AA~ (i.e. excluding 14.75 -- 15.40 \AA) to correct the optical depth effect.}
\tablenotetext{j}{We excluded the central 13 \arcsec\ radius region to eliminate the central non-hot-gas emission.}

\label{tab:1T}

\setlength{\footskip}{-5cm}

\end{deluxetable}

\begin{rotate}

\begin{deluxetable}{l c l l c c c c c c c c c c c c c}
\setlength{\tabcolsep}{2pt}
\tabletypesize{\tiny}
\tablecaption{Two Phase Model Fitting for {\it XMM\/} EPIC and RGS spectra.}
\tablehead{
 \colhead{Galaxy}&  \colhead{Camera/Size} & \colhead{$T1$(keV)} & \colhead{$T2$(keV)} &\colhead{ Fe }  & \colhead{O }& \colhead{Mg} & \colhead{Si} & \colhead{Ne}& \colhead{S}& \colhead{Ni} & \colhead{dof} & \colhead{$\chi^2_r$}& \colhead{Power}& \colhead{Vapec1} & \colhead{Vapec2}&  \colhead{Vapec/Power}\\
}
\startdata

 NGC720 & EPIC 1\arcmin &$0.217_{-0.079}^{+0.161}$ &  $0.567_{-0.017}^{+0.019}$ &  $0.64_{-0.17}^{+0.38}$ & $0.30_{-0.12}^{+0.24}$ & $0.56_{-0.23}^{+0.46}$ & $0.49_{-0.33}^{+0.47}$ & - & - & $1.55_{-0.70}^{+0.69}$ & 196 & 1.116 & 0.54 & 0.10 & 1.62 & 3.18 \\
& EPIC 4\arcmin &$0.238_{-0.043}^{+0.034}$ &  $0.562_{-0.010}^{+0.010}$ &  $1.10_{-0.26}^{+0.11}$ & $0.45_{-0.14}^{+0.22}$ & $0.76_{-0.32}^{+0.73}$ & $0.97_{-0.45}^{+0.93}$ & - & - & - & 500 & 1.094 & 1.48 & 0.66 & 3.65 & 2.92 \\ 
 \hline 
  NGC1399 & EPIC 1\arcmin &$0.793_{-0.004}^{+0.003}$ &  $1.719_{-0.070}^{+0.065}$ &  $1.70_{-0.12}^{+0.08}$ & $0.76_{-0.09}^{+0.07}$ & $1.83_{-0.14}^{+0.11}$ & $1.75_{-0.11}^{+0.09}$ & $1.52_{-0.19}^{+0.19}$ & $1.11_{-0.13}^{+0.12} $ & $4.48_{-0.34}^{+0.34}$ & 773 & 1.263 & 0.39 & 6.76 & 1.17 & 20.57 \\ 
 & EPIC 4\arcmin &$0.809_{-0.002}^{+0.002}$ &  $1.440_{-0.034}^{+0.031}$ &  $1.32_{-0.05}^{+0.05}$ & $0.61_{-0.05}^{+0.05}$ & $1.36_{-0.07}^{+0.07}$ & $1.48_{-0.05}^{+0.05}$ & - & $1.03_{-0.06}^{+0.06}$ & $3.26_{-0.10}^{+0.19}$ & 1367 & 1.193 & 2.33 & 11.38 & 9.76 & 9.08 \\ 
   &RGS 1\arcmin&$0.800_{-0.006}^{+0.013}$ &  1.720 &$1.50_{-0.52}^{+1.41}$ & $1.22_{-0.15}^{+0.32}$ & - & - & - & - & - & 1007 & 1.165 & 0.52 & 12.28 & 0.26 & 24.03 \\ 
 &RGS 3.5\arcmin&$0.660_{-0.048}^{+0.052}$&  $1.092_{-0.025}^{+0.046}$ &$0.91_{-0.09}^{+0.13}$ & $0.63_{-0.08}^{+0.12}$ & -  & $1.49_{-0.28}^{+0.35}$  & - & - & - & 1450 & 1.258 & 3.03 & 5.24 & 21.80 & 8.92 \\ 
 \hline 
  NGC3923 & EPIC 1\arcmin &$0.244_{-0.039}^{+0.043}$ &  $0.566_{-0.010}^{+0.011}$ &  $0.81_{-0.11}^{+0.07}$ & $0.27_{-0.08}^{+0.12}$ & $0.77_{-0.25}^{+0.43}$ & $0.72_{-0.27}^{+0.43}$ & - & $0.20_{-0.10}^{+0.61}$ & - & 313 & 1.186 & 0.61 & 0.65 & 2.85 & 5.74 \\ 
 & EPIC 4\arcmin &$0.247_{-0.025}^{+0.024}$ &  $0.555_{-0.009}^{+0.009}$ &  $0.72_{-0.19}^{+0.25}$ & $0.24_{-0.06}^{+0.09}$ & $0.64_{-0.19}^{+0.30}$ & $0.59_{-0.23}^{+0.32}$ & - & - & - & 589 & 1.024 & 1.72 & 1.35 & 3.93 & 3.06 \\ 
 \hline 
  NGC4406 & EPIC 1\arcmin &$0.630_{-0.023}^{+0.019}$ &  $0.984_{-0.064}^{+0.085}$ &  $0.86_{-0.09}^{+0.11}$ & $0.46_{-0.10}^{+0.12}$ & $0.67_{-0.11}^{+0.12}$ & $0.68_{-0.12}^{+0.13}$ & $0.53_{-0.18}^{+0.22}$ & - & $0.48_{-0.38}^{+0.50}$ & 461 & 1.195 & 0.01 & 0.06 & 0.02 & 10.44 \\ 
 & EPIC 4\arcmin &$0.633_{-0.014}^{+0.012}$ &  $0.938_{-0.024}^{+0.020}$ &  $0.87_{-0.04}^{+0.02}$ & $0.51_{-0.05}^{+0.05}$ & $0.70_{-0.05}^{+0.06}$ & $0.71_{-0.06}^{+0.05}$ & $0.47_{-0.10}^{+0.10}$ & $ 0.59_{-0.09}^{+0.09}$ & $1.94_{-0.24}^{+0.25}$ & 918 & 1.137 & 0.03 & 0.26 & 0.19 & 14.11 \\ 
  &RGS 1\arcmin&  $0.617_{-0.034}^{+0.026}$& 0.984&$0.33_{-0.05}^{+0.03}$ & $0.31_{-0.08}^{+0.10}$ &$0.20_{-0.10}^{+0.09}$  &- & - & - & - & 303 & 1.293 & 0.02 &  0.28 &0.02 & 13.10 \\ 

  &RGS 3.5\arcmin&$0.266_{-0.014}^{+0.029}$ & $0.702_{-0.024}^{+0.023}$ &$0.59_{-0.11}^{+0.11}$ & $0.20_{-0.04}^{+0.04}$ &- & - & - & - & - & 773 & 1.182 & 0.06 &0.20 & 0.62  & 13.42 \\ 

 \hline 
  NGC4472 & EPIC 1\arcmin &$0.734_{-0.005}^{+0.009}$ &  $1.277_{-0.055}^{+0.042}$ &  $1.55_{-0.08}^{+0.09}$ & $0.63_{-0.07}^{+0.08}$ & $1.29_{-0.09}^{+0.10}$ & $1.28_{-0.08}^{+0.09}$ & $0.86_{-0.08}^{+0.16}$ & $0.94_{-0.12}^{+0.12}$ & $1.94_{-0.34}^{+0.17}$ & 719 & 1.375 & 0.32 & 3.72 & 0.74 & 14.12 \\ 
 & EPIC 4\arcmin &$0.773_{-0.006}^{+0.006}$ &  $1.313_{-0.022}^{+0.023}$ &  $1.75_{-0.06}^{+0.07}$ & $0.77_{-0.05}^{+0.06}$ & $1.64_{-0.07}^{+0.08}$ & $1.70_{-0.06}^{+0.07}$ & $1.36_{-0.12}^{+0.12}$ & $1.25_{-0.08}^{+0.08}$ & $3.17_{-0.22}^{+0.23}$ & 1160 & 1.511 & 1.00 & 7.44 & 3.60 & 11.01 \\ 
 &RGS 1\arcmin&$0.682_{-0.019}^{+0.775}$&  $1.021_{-0.294}^{+0.490}$ &$1.13_{-0.28}^{+0.32}$ & $0.83_{-0.24}^{+0.20}$ &- & - & - & - & - & 535 & 1.456 & 0.44 & 5.13 & 2.58 & 17.34 \\ 
 &RGS 3.5\arcmin&$0.656_{-0.055}^{+0.126}$&  $0.967_{-0.534}^{+0.138}$ &$1.86_{-0.38}^{+0.63}$ & $1.31_{-0.25}^{+0.32}$ &$1.38_{-0.34}^{+0.62}$  & -  & - & - & - & 1013 & 1.187 & 1.08 & 3.83&   9.46& 12.35 \\ 

 \hline 
  NGC4552 & EPIC 1\arcmin &$0.261_{-0.139}^{+0.202}$ &  $0.609_{-0.010}^{+0.029}$ &  $0.44_{-0.05}^{+0.10}$ & $0.27_{-0.08}^{+0.10}$ & $0.50_{-0.11}^{+0.20}$ & $0.46_{-0.13}^{+0.20}$ & - & - & - & 413 & 1.056 & 0.07 & 0.01 & 0.18 & 2.66 \\ 
  & EPIC 4\arcmin &$0.353_{-0.059}^{+0.029}$ &  $0.622_{-0.027}^{+0.022}$ &  $0.49_{-0.08}^{+0.08}$ & $0.25_{-0.05}^{+0.07}$ & $0.54_{-0.15}^{+0.23}$ & $0.54_{-0.19}^{+0.26}$ & - & - & - & 628 & 1.132 & 0.12 & 0.09 & 0.15 & 2.00 \\ 

 \hline
   NGC4636 & EPIC 1\arcmin &$0.336_{-0.010}^{+0.013}$ &  $0.606_{-0.009}^{+0.010}$ &  $1.01_{-0.05}^{+0.06}$ & $0.37_{-0.03}^{+0.03}$ & $0.79_{-0.07}^{+0.07}$ & $1.04_{-0.10}^{+0.10}$ & $0.76_{-0.10}^{+0.08}$ & $0.63_{-0.15}^{+0.16}$ & $0.32_{-0.22}^{+0.49}$ & 502 & 1.501 & 0.14 & 1.45 & 3.38 & 33.74 \\ 
 & EPIC 4\arcmin &$0.367_{-0.020}^{+0.018}$ &  $0.644_{-0.006}^{+0.007}$ &  $1.06_{-0.03}^{+0.03}$ & $0.51_{-0.03}^{+0.03}$ & $0.87_{-0.04}^{+0.04}$ & $1.16_{-0.05}^{+0.05}$ & $0.86_{-0.06}^{+0.07}$ & - & $1.44_{-0.28}^{+0.28}$ & 837 & 1.708 & 0.57 & 1.29 & 13.55 & 26.22 \\ 
 &RGS 1\arcmin& 0.336& $0.627_{-0.011}^{+0.012}$ &$0.85_{-0.11}^{+0.14}$ & $0.37_{-0.07}^{+0.08}$ &$0.47_{-0.17}^{+0.20}$  & - & - & - & - & 437 & 2.023 & 0.21 & 0.18 & 6.76 & 32.36 \\ 
 &RGS 3.5\arcmin& $0.272_{-0.145}^{+0.028}$& $0.645_{-0.005}^{+0.009}$ &$1.07_{-0.14}^{+0.14}$ & $0.41_{-0.04}^{+0.14}$ &$0.63_{-0.19}^{+0.18}$  &- & - & - & - & 891 & 1.695 & 0.55 & 0.66 & 13.91 & 26.52 \\ 

 &RGS corrected 1\arcmin& 0.336& $0.586_{-0.007}^{+0.010}$ &$1.08_{-0.14}^{+0.06}$ & $0.55_{-0.06}^{+0.07}$ &$0.62_{-0.14}^{+0.24}$  & -  & - & - & - & 388 & 1.346 & 0.21 & 0.00 & 7.15 & 33.35 \\ 
 &RGS corrected 3.5\arcmin& $0.310_{-0.066}^{+0.029}$& $0.635_{-0.006}^{+0.016}$ &$1.42_{-0.17}^{+0.16}$ & $0.58_{-0.10}^{+0.05}$ &$0.96_{-0.25}^{+0.18}$  &- & - & - & - & 808 & 1.296 & 0.55 & 1.39 & 13.40 & 26.92 \\ 

 \hline 
  NGC4649 & EPIC 1\arcmin &$0.685_{-0.002}^{+0.043}$ &  $0.902_{-0.016}^{+0.105}$ &  $1.70_{-0.13}^{+0.14}$ & $0.57_{-0.09}^{+0.10}$ & $1.69_{-0.15}^{+0.16}$ & $1.68_{-0.14}^{+0.15}$ & $0.69_{-0.25}^{+0.11}$ & $1.24_{-0.20}^{+0.21}$ & $3.70_{-0.50}^{+0.27}$ & 628 & 1.356 & 0.56 & 2.76 & 2.92 & 10.20 \\ 
 & EPIC 4\arcmin &$0.785_{-0.003}^{+0.003}$ &  $1.652_{-0.202}^{+0.292}$ &  $1.62_{-0.10}^{+0.11}$ & $0.63_{-0.08}^{+0.08}$ & $1.51_{-0.11}^{+0.12}$ & $1.72_{-0.11}^{+0.11}$ & $0.85_{-0.09}^{+0.16}$ & $ 1.06_{-0.15}^{+0.15}$ & $2.66_{-0.36}^{+0.18}$ & 930 & 1.421 & 1.53 & 9.14 & 0.73 & 6.45 \\ 

 \hline
   NGC5044 & EPIC 1\arcmin &$0.467_{-0.100}^{+0.177}$ &  $0.768_{-0.009}^{+0.041}$ &  $0.86_{-0.07}^{+0.07}$ & $0.41_{-0.08}^{+0.10}$ & $0.84_{-0.10}^{+0.11}$ & $0.86_{-0.09}^{+0.10}$ & $0.33_{-0.18}^{+0.18}$ & $0.76_{-0.16}^{+0.16}$ & $2.41_{-0.44}^{+0.23}$ & 474 & 1.074 & 1.73 & 46.70 & 2.45 & 28.33 \\  
 & EPIC 4\arcmin &$0.786_{-0.003}^{+0.003}$ &  $1.474_{-0.038}^{+0.051}$ &  $1.15_{-0.05}^{+0.05}$ & $0.49_{-0.05}^{+0.05}$ & $1.14_{-0.06}^{+0.06}$ & $1.01_{-0.05}^{+0.02}$ & $ 0.80_{-0.10}^{+0.10}$ & $0.81_{-0.07}^{+0.07} $ & $2.11_{-0.18}^{+0.19}$ & 830 & 1.129 & 1.78 & 206.98 & 49.24 & 144.21 \\ 
 &RGS 1\arcmin&$0.765_{-0.015}^{+0.015}$&  1.474 &$0.77_{-0.12}^{+0.14}$ & $0.48_{-0.11}^{+0.12}$ &- & - & - & - & - & 295 & 1.257 & 4.40 & 114.10& 10.12  & 28.23 \\ 
 &RGS 3.5\arcmin&$0.671_{-0.038}^{+0.132}$& $0.908_{-0.049}^{+0.481}$  &$0.64_{-0.06}^{+0.06}$ & $0.46_{-0.07}^{+0.07}$ &$0.49_{-0.14}^{+0.14}$  & $0.87_{-0.18}^{+0.21}$  & - & - & - & 666 & 1.202 & 2.10 &  115.57&  186.75& 144.31 \\ 

 \hline 
 
\enddata

NOTE:~Table is the same as previous one, only that we use two phase model here: {\it wabs(vapec+vapec+power)} for EPIC, and {\it wabs(vapec+vapec+power)*rgsxsrc} for RGS. For RGS spectra, the second temperature will be frozen to the EPIC fits if it cannot be constrained.
\label{tab:2T}

\end{deluxetable}
\end{rotate}

\begin{figure}
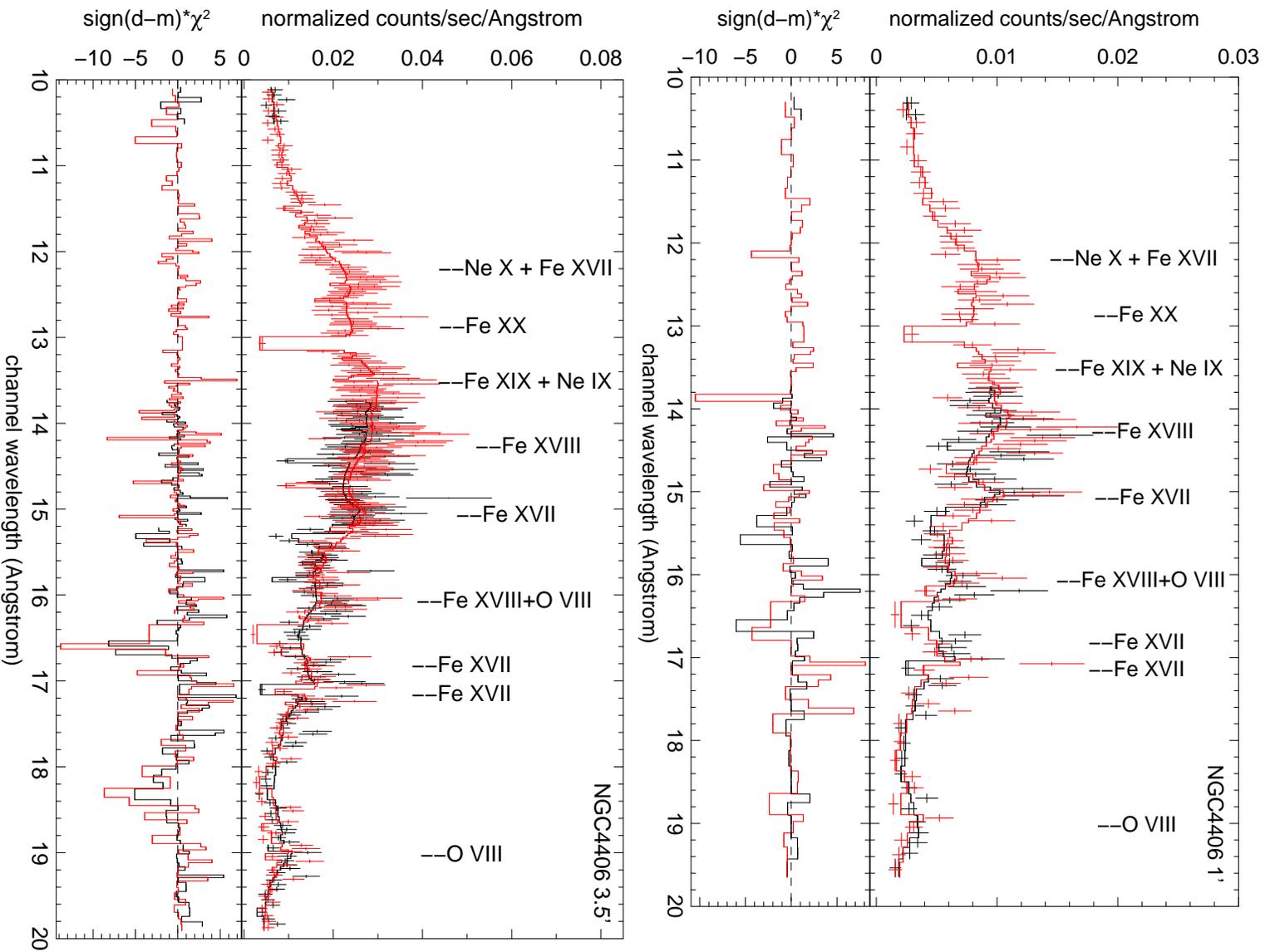

\epsscale{1.0}
\includegraphics[width=4in,angle=-90 ]{f5a.eps}
\includegraphics[width=4in,angle=-90 ]{f5b.eps}
\caption[RGS:n4406]%
        {Same as Fig.~\ref{fig:n1399}, but for NGC 4406.}
\end{figure}

\begin{figure}
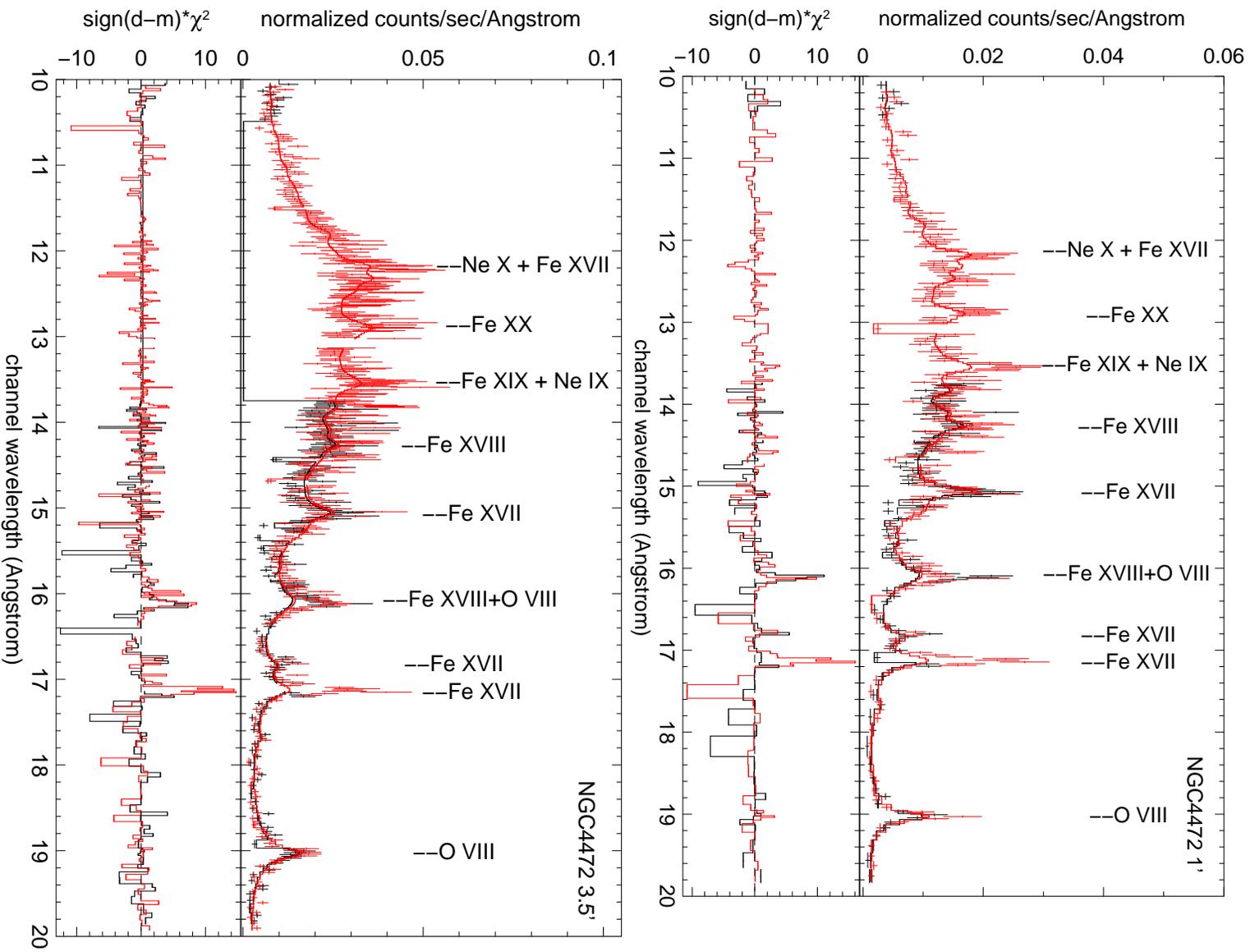

\epsscale{1.0}
\includegraphics[width=4in,angle=-90 ]{f6a.eps}
\includegraphics[width=4in,angle=-90 ]{f6b.eps}
\caption[RGS:n4472]%
        {Same as Fig.~\ref{fig:n1399}, but for NGC 4472.}
\end{figure}

\begin{figure}
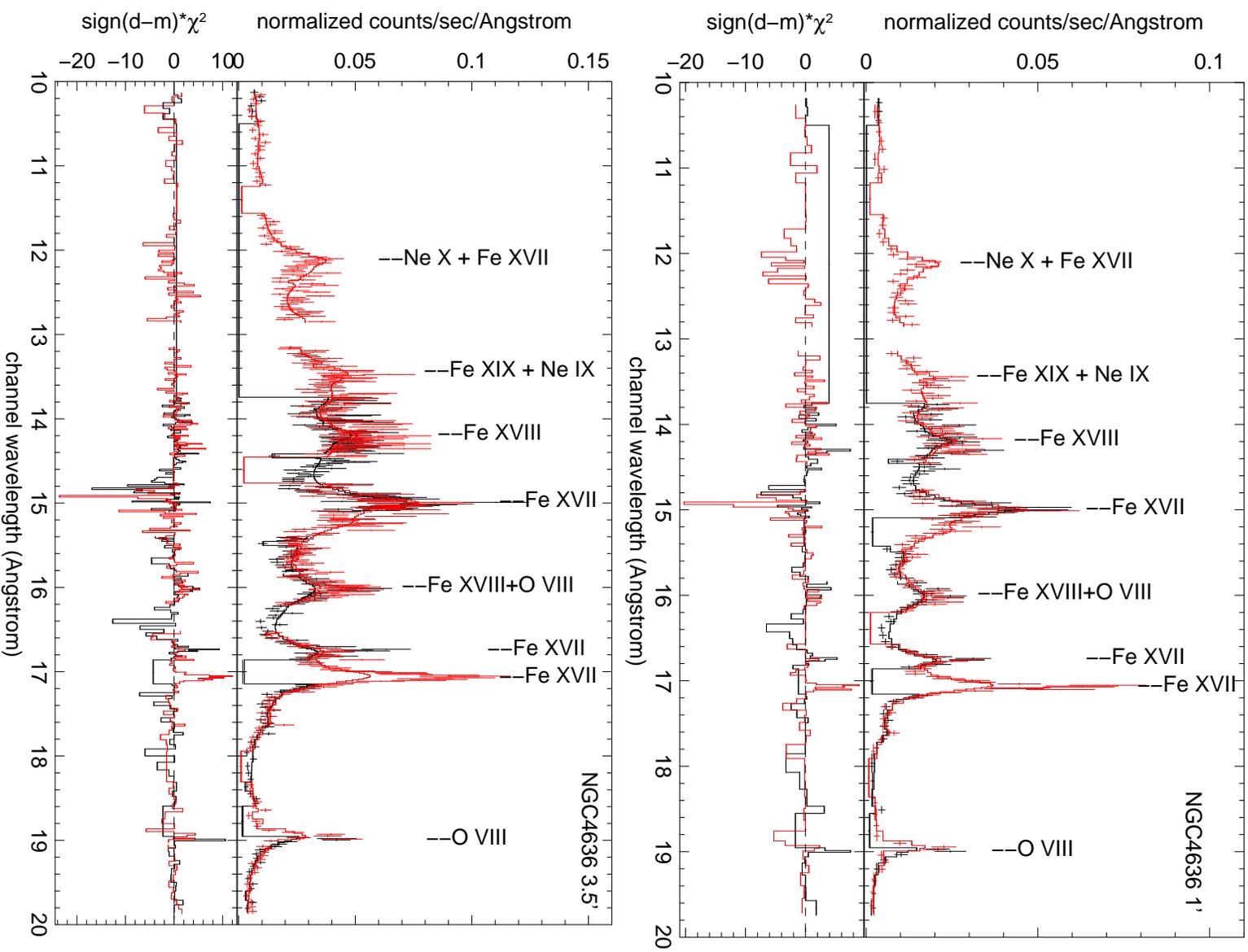

\epsscale{1.0}
\includegraphics[width=4in,angle=-90 ]{f7a.eps}
\includegraphics[width=4in,angle=-90 ]{f7b.eps}
\caption[RGS:n4636]%
        {Same as Fig.~\ref{fig:n1399}, but for NGC 4636.}
\end{figure}

\begin{figure}
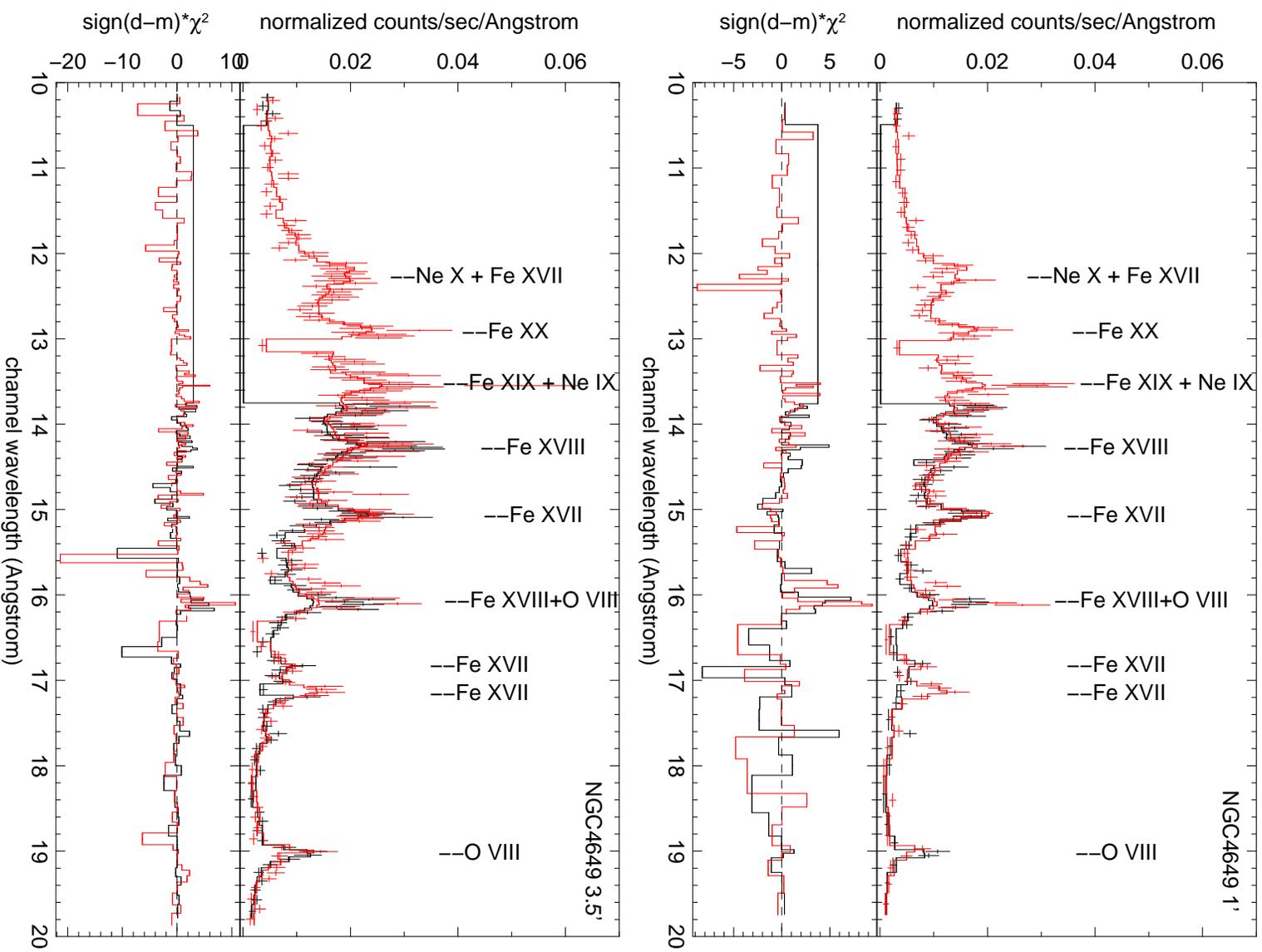

\epsscale{1.0}
\includegraphics[width=4in,angle=-90 ]{f8a.eps}
\includegraphics[width=4in,angle=-90 ]{f8b.eps}
\caption[RGS:n4649]%
        {Same as Fig.~\ref{fig:n1399}, but for NGC 4649.}
\end{figure}

\begin{thebibliography}{}


\bibitem[Anders \& Grevesse(1989)]{ande89} Anders, E., \& 
Grevesse, N.\ 1989, \gca, 53, 197 

\bibitem[Arimoto et al.(1997)]{arim97} Arimoto, N., 
Matsushita, K., Ishimaru, Y., Ohashi, T., \& Renzini, A.\ 1997, \apj, 477, 
128 

\bibitem[Athey(2007)]{athey07} Athey, A.~E.\ 2007, Ph.D.~Thesis,arXiv:0711.0395v1

\bibitem[Brown \& Bregman(1998)]{brown98} Brown, B.~A., \& Bregman, J.~N. \ 1998, \apjl, 495,L75 

\bibitem[Buote et al.(1998)]{buote98} Buote, D.~A., \& Fabian,A.~C.\ 1998, \mnras, 296,977 

\bibitem[Buote(1999)]{buote99} Buote, D.~A.\ 1999, \mnras, 309,685 

\bibitem[Buote(2000)]{buote2000} Buote, D.~A.\ 2000, \mnras, 311, 
176 

\bibitem[Buote(2002)]{buote02} Buote, D.~A.\ 2002, \apjl,574,L135

\bibitem[Buote et al.(2003)]{buote03} Buote, D.~A.,Lewis, A.~D., Brighenti,F., \& Mathews, W. \ 2003, \apj, 595, 
151

\bibitem[De Vaucouleurs et al.(1991)]{devauc91} de Vaucouleurs, G., de Vaucouleurs, A., Corwin, H.~G., Buta, R.~J., Paturel, G., \& Fouque, P. \ 1991, Third Reference Catalogue of Bright Galaxies (Volume 1-3, XII, 2069 pp. 7 figs.. Springer-Verlag Berlin Heidelberg New York) 

\bibitem[Diehl \& Statler(2007)]{diehl2007} Diehl, S., \& Statler, T.~S.\ 
2007, ArXiv e-prints, 711, arXiv:0711.3071 

\bibitem[Dickey \& Lockman(1990)]{dick90} Dickey, J.~M., \& 
Lockman, F.~J.\ 1990, \araa, 28, 215 

\bibitem[Humphrey \& Buote(2006)]{hump06} Humphrey, P.~J., \& 
Buote, D.~A.\ 2006, \apj, 639, 136 

\bibitem[Fabbiano et al.(2003)]{Fabbiano03} Fabbiano,G., Elvis,M., Markoff,S., Siemiginowska, A., Pellegrini,S.,
 Zezas,A., Nicastro,F., Trinchieri,G., \& McDowell,J.\ 2003, \apj, 588, 175

\bibitem[Fulbright et al.(2004)]{Fulbright04} Fulbright, J.~P., Rich, R.M., \& Mcwilliam, A.\ 2005, Nucl. Phys. A, 758, 197 

\bibitem[Gastaldello \& Molendi(2002)]{2002ApJ...572..160G} Gastaldello, 
F., \& Molendi, S.\ 2002, \apj, 572, 160 

\bibitem[Grevesse \& Sauval(1998)]{grev98} Grevesse, N., \& 
Sauval, A.~J.\ 1998, Space Science Reviews, 85, 161 

\bibitem[Irwin et al.(2003)]{irwin03} Irwin, J.~A., Athey, 
A.~E., \& Bregman, J.~N.\ 2003, \apj, 587, 356 

\bibitem[Irwin et al.(2002)]{irwin02} Irwin, J.~A., Sarazin, 
C.~L., \& Bregman, J.~N.\ 2002, \apj, 570, 152 

\bibitem[Jones et al.(2002)]{jones2002} Jones, C., Forman, W., 
Vikhlinin, A., Markevitch, M., David, L., Warmflash, A., Murray, S., 
\& Nulsen, P.~E.~J.\ 2002, \apjl, 567, L115

\bibitem[Kawata \& Gibson(2003)]{2003MNRAS.346..135K} Kawata, D., \& 
Gibson, B.~K.\ 2003, \mnras, 346, 135 

\bibitem[Kim \& Fabbiano(2003)]{kim03} Kim, D.-W., \& 
Fabbiano, G.\ 2003, \apj, 586, 826 

\bibitem[Kim \& Fabbiano(2004)]{kim04} Kim, D.-W., \& 
Fabbiano, G.\ 2004, \apj, 613, 933

\bibitem[Minniti \& Zoccali(2007)]{Minniti07} Minniti, D. \& Zoccali, M. \ 2007, 
ArXiv e-prints, 710, arXiv:0710.3104v1

\bibitem[Machacek et al.(2006)]{machacek06} Machacek, M., Jones, C., Forman, W. R., \& Nulsen, P. \ 2006, 
\apj, 644, 155 

\bibitem[Nomoto et al.(1997)]{nomoto97} Nomoto, K., Iwamoto, K., Nakasato, N., Thielemann, F. K., Brachwitz, F., Tsujimoto, T., Kubo, Y., \& Kishimoto, N. \ 1997, Nucl. Phys. A, 621, 467

\bibitem[O'Sullivan \& Ponman(2004)]{osul04} O'Sullivan, E., 
\& Ponman, T.~J.\ 2004, \mnras, 349, 535 

\bibitem[O'Sullivan et al.(2005)]{osul05} O'Sullivan, E., 
Vrtilek, J.~M., \& Kempner, J.~C.\ 2005, \apjl, 624, L77 

\bibitem[Park et al.(2003)]{park03} Park, S., Hughes, J.~P., Slane, P.~O., 
Burrows, D.~N., Warren, J.~S., Garmire, G.~P., \& Nousek, J.~A. \ 2003, \apj, 592, L41 

\bibitem[Randall et al.(2004)]{rand04} Randall, S.~W., 
Sarazin, C.~L., \& Irwin, J.~A.\ 2004, \apj, 600, 729 

\bibitem[Randall et al.(2006)]{rand06} Randall, S.~W., 
Sarazin, C.~L., \& Irwin, J.~A.\ 2006, \apj, 636, 200 

\bibitem[Sambruna et al.(2004)]{samb04} Sambruna, R.~M., 
Gliozzi, M., Donato, D., Tavecchio, F., Cheung, C.~C., \& Mushotzky, R.~F.\ 
2004, \aap, 414, 885 

\bibitem[Smith et al.(2001)]{smith01} Smith, R.~K., Brickhouse, 
N.~S., Liedahl, D.~A., \& Raymond, J.~C.\ 2001, \apjl, 556, L91 

\bibitem[Stickel et al.(2003)]{stickel2003} Stickel, M., Bregman, J.~N., 
Fabian, A.~C., White, D.~A., \& Elmegreen, D.~M.\ 2003, \aap, 397, 503 

\bibitem[Tamura et al.(2003)]{tamu03} Tamura, T., Kaastra, 
J.~S., Makishima, K., \& Takahashi, I.\ 2003, \aap, 399, 497 

\bibitem[Tawara et al.(2008)]{tawara08} Tawara, Y., Matsumoto, C., Tozuka, M., Fukazawa, Y., Matsushita, K., \& Anabuki,N.\ 2008, \pasj, 60, S307 

\bibitem[Tonry et al.(2001)]{tonry01} Tonry, J. ~L, Dressler, A., Blakeslee, J. ~P., Ajhar, E. ~A., Fletcher, A. ~B., Luppino, G. ~A., \& Moore,C.~B. \ 2001, \apj, 546, 681 

\bibitem[Trager et al.(2000)]{trager00} Trager, S. ~C., Faber, S. ~M., Worthey, G., \& Gonzalez, J. ~J. \ 2000, \aj, 120, 165

\bibitem[Trinchieri et al.(1994)]{trinc94} Trinchieri, G., Kim, 
D.-W., Fabbiano, G., \& Canizares, C.~R.~C.\ 1994, \apj, 428, 555 


\bibitem[Xu et al.(2002)]{xu02} Xu, H., et al.\ 2002, \apj, 
579, 600 

\end{thebibliography}
\end{document}